\documentclass[twocolumn,showpacs,prb,superscriptaddress,aps,floatfix]{revtex4-1}

\usepackage{graphicx}% Include figure files
\usepackage{dcolumn}% Align table columns on decimal point
\usepackage{bm}% bold math
\usepackage{nicefrac}
\usepackage{hyperref}% add hypertext capabilities
\usepackage{gensymb}  %to get deg symbol for example
\usepackage{amsmath}
\usepackage{xcolor}
\usepackage{amssymb}
\usepackage{soul}
\usepackage{float}

\def\kk         {{\bf k}}
\def\qq		{{\bf q}}

\newcommand{\cinam}{Aix-Marseille Univ., CNRS, CINaM, Centre Interdisciplinaire de Nanoscience de Marseille, UMR 7325, Campus de Luminy, 13288 Marseille cedex 9, France}
\newcommand{\piim}{Aix Marseille Univ., PIIM, Physique des Interactions Ioniques et Moléculaires, UMR 7345, 13397, Marseille, France}
\newcommand{\etsf}{European Theoretical Spectroscopy Facility (ETSF)}

\newcommand*{\CA}[1]{\textcolor{magenta}{[CA: #1]}}

\graphicspath{{./}}

\begin{document}

\title{Pressure Dependence of Electronic, Vibrational and Optical Properties of wurtzite-Boron Nitride}% Force line breaks with \\

\author{Martino Silvetti}
\email{martino.silvetti@univ-amu.fr}
\affiliation{\piim}
\affiliation{\etsf}
\author{Claudio Attaccalite}%
\affiliation{\cinam}
\affiliation{\etsf}
\author{Elena Cannuccia}
\affiliation{\piim}
\affiliation{\etsf}
\date{\today}% It is always \today, today,%  but any date may be explicitly specified

\begin{abstract}
Wurtzite Boron Nitride ($w$BN) is a wide band gap BN polymorph with unique mechanical properties such as hardness and stiffness. Initially synthesized in 1963 by transforming hexagonal BN ($h$BN) under high temperature and pressure conditions, $w$BN can now be stabilised at atmospheric pressure to obtain high-quality samples. Our first-principles study investigates the electronic, vibrational and optical properties of $w$BN across a broad range of pressures. We account for  the electron-hole interaction in the optical response, revealing that this effect is crucial to interpret the available experimental spectra. We also calculate the coupling between excitons and phonons and provide for the first time a phonon-assisted emission spectrum, centered around 6.02~eV. Our results hold significant importance for the potential application of $w$BN as a dielectric material in BN-based technologies, especially in optoelectronics and harsh environments. We also expect that our prediction could be verified in the future, and it could aid in the identification of $w$BN through cathodoluminescence experiments.   
\end{abstract}

%\keywords{Suggested keywords}%Use showkeys class option if keyword display desired
\maketitle

%\tableofcontents
%======================INTRO===========
\section{\label{sec:intro}Introduction}
%======================================
Recent advances in high pressure physics have enabled the synthesis of a new range of materials that were previously impossible to create at ambient pressure.\cite{RevModPhys.90.015007} Pressure has the ability to drastically alter elastic, electronic, magnetic, structural and chemical properties, transforming materials from simple insulators to superconductors, from amorphous to crystalline solids, and from ionic to covalent compounds. Some of these transformations can be irreversible, such as the transition from graphite to diamond, resulting in the creation of new materials that are stable at ambient pressure.

Among the various materials created by the use of pressure are wide band gap insulators.\cite{segura2021tuning} The term "wide band gap material" generally refers to any semiconductor with an energy band gap much larger than that of conventional semiconductors such as silicon (Si) and gallium arsenide (GaAs). Interest in this class of materials stems from the ongoing quest for smaller, faster, more reliable and more efficient electronic devices than their Si-based counterparts. Wide band gap materials have promising applications for future generations of high power electronics, deep UV optoelectronics, quantum electronics and harsh environment (high temperature and high pressure) applications.

%Wide-bandgap semiconductor technologies have actively been sought as the next generation semiconductors due to their inherent wide bandgap, high electron saturation velocity, high thermal conductivity, and wide temperature operation range capabilities. For example, a wide bandgap semiconductor can sustain high temperature operation up to about 300 C. Compare this with a maximum temperature of roughly 100 C for silicon-based microelectronics. This property will enable microelectronic systems to more easily operate in elevated ambient temperature environments such as power plants and hybrid electric vehicles (HEVs). Additionally, the critical electric field for wide-bandgap materials is typically ten times higher than that for any current commercially available semiconductor technology. This feature will allow wide-bandgap electronic circuits to operate at very high voltages with high power amplification efficiency. 

Wide band gap materials are the optimal choice for efficient green and blue light emitting diodes (LEDs) in optoelectronic applications. For example, gallium nitride (GaN) is used to produce bright blue LEDs.\cite{akasaki2014nobel} At shorter wavelengths, these materials find applications in UV light germicidal sources for water sanitation and advanced solar blind optical communication systems. Recently, hexagonal boron nitride ($h$BN), a wide band gap material, has attracted much attention for its very efficient UV light emission despite being an indirect band gap material.\cite{Watanabe2004,Cassabois2016} This discovery has led to a number of theoretical and experimental studies focusing on its unexpected properties.\cite{Schue2016,Fossard2017,vuong2018,Sponza2018,Schue2019,Cannuccia2019,Paleari2019,Roux2021}

However, compared to currently commercialised semiconductor technologies based on Si, GaAs and even indium phosphide (InP), applications of wide band gap materials are relatively immature. Nevertheless, significant research and development efforts are underway due to the enormous commercialisation opportunities for wide-bandgap semiconductors in next-generation microelectronics and optoelectronics. In some cases, much remains to be done to accurately characterise their electronic and optical properties.

In this context, wurtzite boron nitride ($w$BN), which is isostructural to other group III nitrides with optoelectronics applications such as GaN, AlN and InN, stands out due to its exceptional mechanical properties such as hardness and elastic stiffness\,\cite{Nagakubo2013,Deura2017} and its electronic properties such as high thermal conductivity and large spontaneous polarisation\,\cite{Dreyer2014,Yixi1994}.

Since its first synthesis in 1963, much effort has been devoted to the production of high quality samples of wurtzite boron nitride ($w$BN), which is obtained by subjecting hexagonal boron nitride ($h$BN) to high temperature and pressure conditions. For some time, the dominant synthesis method has been the shock compression technique of $h$BN, but unfortunately this only produces $w$BN powders with grain sizes in the $\mu$m range. However, in 2019, researchers successfully synthesised pure single-phase polycrystalline $w$BN bulk material using $w$BN powders as a starting point and a well-controlled process under ultra-high pressure (200 kBar) and high temperature (1150\degree C) conditions\,\onlinecite{Liu2019}. The main technical obstacle in the $h$BN$\rightarrow$$w$BN transformation is the stabilisation of the latter at atmospheric pressure, since it tends to revert to the ambient phase due to its low kinetic barrier. Several efforts have been made to overcome this barrier and the wurtzite phase has been stabilised at atmospheric pressure. Finally, $w$BN has recently been obtained from $h$BN in an upstroke cycle from ambient pressure conditions up to 200 kBar.

Optical measurements of wurtzite boron nitride ($w$BN) in the IR-visible spectral energy range have shown that the hexagonal to wurtzite phase transition is complete at 130 kBar\,\cite{Segura2019-nonreversiblewbn} and the optical features of the $w$BN structure are retained in a pressure down-stroke cycle back to ambient conditions. In addition, Chen et al.\cite{Chen2019} successfully fabricated millimeter-sized $w$BN crystals via high-pressure, high-temperature transformation and demonstrated that the stability of $w$BN at atmospheric pressure is guaranteed by the presence of a 3D high-density network of planar defects.

Motivated by recent experimental progress in obtaining high-quality crystalline samples of $w$BN and the pressure-induced phase transition from $h$BN to $w$BN, we present here the first theoretical study of the optical properties (light absorption and emission) of $w$BN in the UV energy range over a wide pressure range from ambient to 200 kBar. We also investigate the electronic and vibrational properties of $w$BN as a function of pressure to support the discussion. Our aim is to extend the understanding of the intrinsic light absorption and emission properties of $w$BN and to complement ongoing research on the cubic phase of boron nitride ($c$BN), a wide band gap material\,\cite{Tararan2018}, which is commonly synthesised at high pressure and high temperature using a temperature gradient\,\cite{Mishima2000}. It is worth noting that $w$BN is actually an intermediate product of the phase transformation of $h$BN to $c$BN and that the synthesis of high purity $c$BN is hindered by the presence of various types of defects, including point defects and stacking defects, in the form of very local $w$BN phase domains\,\cite{Horiuchi1996}. This hypothesis has been proposed due to the existence in cathodoluminescence spectra of phonon replicas incompatible with $c$BN phonon modes, which on the other hand could be the signal for the presence of $w$BN domains\,\cite{Tararan2018}. For these reasons, theoretical reference spectra of pure $w$BN are an important contribution and will allow to elucidate the above mentioned aspects of the phase transformation.

The paper is organized as follows. In Section\,\ref{sec:theo} we give an overview of the theoretical methods used. In Section\,\ref{sec:comp} we present all the computational details necessary to reproduce the results discussed in Section\,\ref{sec:results}, where we investigate the vibrational, electronic and optical properties of $w$BN under hydrostatic pressure. We address the issue of the electronic and optical band gap and compare it with that of other boron nitride phases. Finally, we discuss the phonon-assisted light emission properties of $w$BN and speculate on possible changes in the light emission behaviour with increasing pressure.

%======================================
\section{Theoretical methods}
\label{sec:theo}
%======================================
Using Density Functional Theory (DFT) as implemented in the Quantum Espresso package\,\cite{qe-code}, we investigated the ground state of the wurtzite phase of boron nitride. To optimize the unit cell and atomic positions at various pressures, we minimized the total enthalpy. Subsequently, we utilized Density Functional Perturbation Theory (DFPT)\,\cite{Giannozzi_2017} to calculate the vibrational properties based on the optimized atomic structures. These methods allowed us to gain a comprehensive understanding of the structural and vibrational characteristics of the boron nitride wurtzite phase. We used PBEsol exchange-correlation functional\,\onlinecite{perdew2008restoring} and DOJO pseudopotentials.\onlinecite{van2018pseudodojo} Technical details of these calculations are given in the following section.

Following our investigation of the ground-state properties, we shifted our focus to exploring the electronic structure and optical properties of $w$BN. To study the electronic bands of $w$BN, we diagonalized the Kohn-Sham (KS) Hamiltonian. However, as DFT is known to have limitations in accurately describing the electronic structure, including underestimation of the electronic gap,\cite{onida_reining_rubio} we used Many-Body Perturbation Theory implemented in the Yambo code\,\cite{yambo-code} to correct the KS eigenvalues. By doing so, we calculated the quasi-particle band structure as:
\begin{equation}
	E_{n\kk}=\epsilon_{n\kk}^{KS}  + Z_{n \kk} \left [Re \Sigma(\epsilon_{n\kk}^{KS}) - V^{xc}_{n\kk} \right ]
\end{equation}
where $E_{n\kk}$ are the corrected quasi-particle energies, $Z_{n \kk}$ is the renormalization factor, $V^{xc}$ the exchange correlation KS potential and the self-energy $\Sigma$ is calculated in the GW approximation $\Sigma = G_0W_0$ \cite{onida_reining_rubio}, at frequency $\omega=\epsilon_{n\kk}$. 

Starting from the quasi-particle band structure we proceeded to calculate the optical response function, including the electron-hole interaction. To achieve this, we solved the Bethe-Salpeter Equation (BSE)\,\cite{strinati}, which provides a means of describing the behavior of interacting electrons and holes. The BSE can be reformulated as an eigenvalue problem, which is determined by the two-particle Hamiltonian\,\cite{onida_reining_rubio} in the following manner:
\begin{eqnarray}
	\begin{split}
\label{eq:H_BSE}
H_{\substack{vc\boldsymbol{k}\\v'c'\boldsymbol{k}'}}=\left(E_{c\boldsymbol{k}}-E_{v\boldsymbol{k}}\right)\delta_{vv'}\delta_{cc'}\delta_{\boldsymbol{k}\boldsymbol{k}'}+\\+\left(f_{c\boldsymbol{k}}-f_{v\boldsymbol{k}} \right)\left(2\overline{V}_{\substack{vc\kk\\v'c'\kk'}} - W_{\substack{vc\boldsymbol{k}\\v'c'\boldsymbol{k}'}}\right)
	\end{split}
\end{eqnarray}
where $E_{n\boldsymbol{k}}$ and $f_{n\boldsymbol{k}}$ (where $n$ runs over the valence ($v$) and conduction bands ($c$)) are the quasi-particle energies and occupations respectively, $\overline{V}$ is the Coulomb potential derived from the variation of the Hartree term and $W$ the screened electron-hole interaction derived from the screened exchange.\cite{strinati}
After diagonalization of Eq.~\ref{eq:H_BSE} we obtain the eigenvectors \(A^{S}_{cv\boldsymbol{k}}\) and eigenvalues \(E_{S} \) that are the eigenstates and the energies of the excitons (labelled by the index $S$), the stationary state of the BSE. These quantities are used to build up the macroscopic dielectric function \(\epsilon_{M}\) as:
\begin{equation}
\epsilon_M(\omega)= 1 - 4\pi\sum_S \frac{ | T_S|^2}{\omega - E_S + i\eta},
\label{eq:epsilon}
\end{equation}
where $T_S =  \sum_{cv\kk} d_{cv\kk} A^S_{cv\kk}$ are the excitonic dipoles, $d_{cv\kk}$ are the dipole matrix elements between the Kohn-Sham states $d_{cv\kk} = \langle v \kk | \hat r | c\kk \rangle$ and $i\eta$ is a small broadening term added to simulate experimental spectra.

In addition to studying the optical absorption properties of $w$BN, we were also interested in investigating its luminescence behavior. However, since $w$BN is an indirect band gap material, the emission of light is only possible if it is assisted by phonon modes, as has been observed in $h$BN\,\cite{Cassabois2016}. Thus, it is essential to consider the exciton-phonon coupling when analyzing luminescence in $w$BN. There are two approaches to account for this coupling: the finite differences approach\,\cite{Paleari2019,Cannuccia2019}, and a direct calculation of the exciton-phonon dipole matrix elements\,\cite{Cheng2020}. In this manuscript, we adopted the former method, and the calculation details are provided in the following section. After obtaining the second derivatives of the exciton dipoles, we proceeded with the calculation of the luminescence spectra using a generalization of the Roosbroeck-Shockley relation\,\cite{Roosbroeck1954} for the excitonic case, as derived in Refs.~\cite{Paleari2019,Bebb1972}:
\begin{eqnarray}
\label{eq:RS}
        R^{\mathrm{sp}}(\omega)= \sum_{\lambda\qq} \frac{\omega(\omega + 2\Omega_{\lambda\qq})^2}{\pi^2 \hbar c^3 } n_r(\omega) \sum_S \frac{\partial^2 |T_S|^2 }{\partial R_{\lambda\qq}^2}\Bigr|_{\mathrm{eq}} \\
        \Im \left\{\frac{1}{\hbar\omega-(E_S-\Omega_{\lambda\qq})+\mathrm{i}\eta}\right\} B \left (E_S,T_{exc} \right),\nonumber
\end{eqnarray}
where $\lambda,\qq$ are the phonon mode indexes, $n_r(\omega)$ is the refractive index, $B \left (E_S,T_{exc} \right)$ is the Boltzmann occupation of the lowest excitons and $\Omega_{\lambda \qq}$ are the phonon frequencies. The formula we used provides a luminescence spectrum that accounts for the coupling with phonon modes under equilibrium conditions, where the excitons have relaxed at the minimum of their band structure. 

%=======================================
\section{Computational details}
\label{sec:comp}
\begin{figure}
\includegraphics[width=0.45\textwidth]{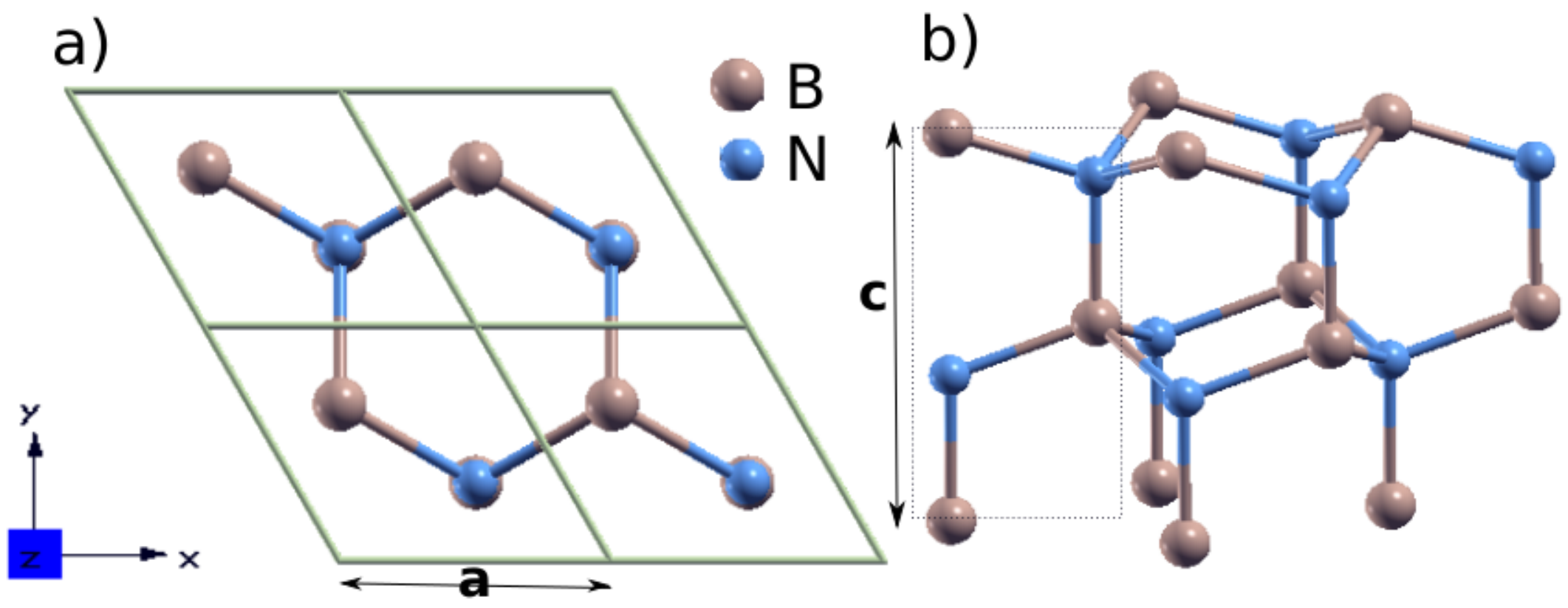}
	\caption{Top view a) and three dimensional structure b) of wurtzite BN. The atoms in the unit cell are framed by a dotted line in panel b), lattice parameter $c$ is parallel to the z-axis. }
	\label{fig:unit_cell}
\end{figure}

%======================================
%\subsection*{\label{subsec:crystal_structure} Crystal Structure}
In this section, we present all the necessary computational details to reproduce the results of our work using the theoretical methods outlined in the previous section.
\emph{Atomic structure}\\
The wurtzite phase has hexagonal symmetry. The crystal structure can be seen as two interpenetrating hexagonal close-packed (hcp) sub-lattices, 
each of them formed by one atom type. The two individual atom types are displaced along the c-axis of amount $z$ in a ABAB... stacking sequence. 
Boron atoms sit in the $(\nicefrac{1}{3},\nicefrac{2}{3},0)$, $(\nicefrac{2}{3},\nicefrac{1}{3},\nicefrac{1}{2})$ crystallographic positions, while nitrogen atoms in $(\nicefrac{1}{3},\nicefrac{2}{3},z)$, $(\nicefrac{2}{3},\nicefrac{1}{3},\nicefrac{1}{2}+z)$ ones.    
The unit cell is shown in Fig.~\ref{fig:unit_cell}. We used the lattice parameters $\bm{a}=2.55$ \AA, $\bm{c}=4.22$ \AA\, taken from Ref.~\onlinecite{Izyumskaya2017} as a starting point parameters to relax the structure at the different pressures. The internal parameter $z$ is defined as the length of the BN bond along the z-axis (in the present case $1.56$ \AA) in units of the $\bm c$ lattice parameter.

We sampled the pressure range 0-200 kBar at a pace of 50 kBar.
Then we relaxed the lattice parameters and the atomic positions at each pressure by sampling the Brillouin zone with a 12$\times$12$\times$8 $\kk$-grid and 70 Ry for the kinetic energy cutoff of the plane wave expansion. The forces acting on the cell and on the atoms were converged to be lower than $10^{-5}$ a.u. Lattice parameters as a function of pressure are summarized in SM\,\cite{SM}, Fig. S1 and Table S1.   
We used the final coordinates and lattice parameters as starting points for investigating the vibrational, electronic and optical properties of $w$BN, that we are going to discuss in the following. \\

%======================================
\emph{Vibrational Properties}\\
%======================================
Starting from the atomic geometries obtained for each pressure chosen in the range 0-200 kBar, the vibrational properties (phonon frequencies, eigenvectors, dielectric constants and Born effective charges) of $w$BN as a function of pressure have been calculated. The dynamical matrices have been sampled on a regular 9$\times$9$\times$6 \(\boldsymbol{q}\)-grid. Then a Fourier interpolation is used to obtain the entire phonon dispersion along a \(\boldsymbol{q}\)-path connecting high symmetry points of the Brillouin zone. Being 4 atoms in the unit cell, 12 phonon modes appear for each $\qq$-point.\\

%======================================
\emph{Electronic band structure and quasi-particle corrections}\\
%======================================
The quasi-particle band structure was obtained starting from the KS Hamiltonian and correcting the eigenvalues within $G_0W_0$ approximation\cite{aryasetiawan1998gw}. We used a \(\boldsymbol{k}\)-grid 12$\times$12$\times$8 k-points grid, and 75 bands for the expansion of G and W, a cutoff of 3 Ha for the dielectric constant and the Godby-Needs plasmon-pole model\cite{PhysRevLett.62.1169} for the dynamical part of W.\\

%======================================
\emph{Light Absorption}\\
%======================================
Optical properties of $w$BN have been investigated by solving the Bethe-Salpeter equation as explained in previous section.\cite{strinati} We used a 12$\times$12$\times$8 k-points grid and included 3 valence and 3 conduction bands and the same screened Coulomb interaction of the GW calculations.\\ 

%======================================
\emph{Phonon assisted luminescence} \\
%======================================
\begin{figure}
	\centering
	\includegraphics[width=0.4\textwidth]{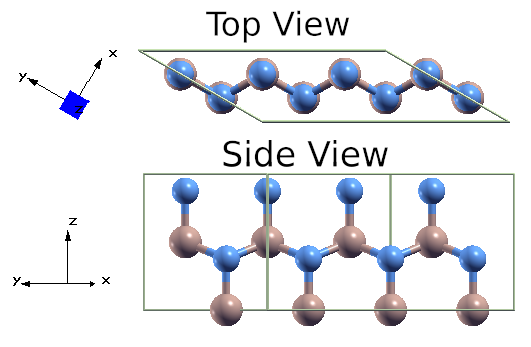}
	\caption{Top view and side view of the supercell mapping the $K$ point at $\Gamma$. }
	\label{fig:331supercell}
\end{figure}
As it will be discussed in the Sec.~\ref{sec:results} $w$BN is an indirect gap material with the lowest indirect gap occurring between points $\Gamma$ and $K$. Therefore the momentum responsible for the indirect emission would be then $\mathbf{q}=K-\Gamma=K=(\nicefrac{1}{3},\nicefrac{1}{3},0)$. In order to get the phonon-assisted luminescence spectra we apply the following strategy: 
\begin{enumerate}

\item We diagonalized the Bethe Salpeter equation at the transferred momentum $\mathbf{q}=K$, first including and then discarding the quasi particle corrections. In this way we estimated a scissor correction that exactly reproduces the position of the exciton at $K$. In Table \ref{tab:IND_excitons_values} we report the position of the exciton at $\qq = K$, with and without GW corrections, and the obtained energy shift. 
%summarize the eigenvalues obtained. We reported just the first two excitons (degenerate) below the KS and QP band gap values, 5.01 and 6.42 eV respectively.
\begin{table}[h]
	\centering
	\begin{tabular}{|c|c|c|}
	\hline 
	\multicolumn{3}{|c|}{Lowest exciton at $\mathbf{q}=K$} \\ 
		\hline
		Exc. En. (IP) &  Exc. En. (GW) & Est. scissor corr.  \\
		    (eV)     &      (eV)             &   (eV) \\
	\hline
	            4.73     &       6.14            &  1.41\\
	\hline
	\end{tabular}
	\caption{Lowest indirect excitons energies at KS+BSE ($1^{st}$ column) and GW+BSE level ($2^{nd}$ column) and the estimated scissor correction ($3^{rd}$ column).\label{tab:IND_excitons_values}}
\end{table}

\item We built a supercell which maps the point $K$ at $\Gamma$ (see Fig.\,\ref{fig:331supercell})\cite{lloyd2015lattice} and we calculated the dielectric constant that enters in BSE with the parameters of the primitive cell (polarization function number of bands) rescaled to the supercell size. Then  we solved the BSE on the supercell at transferred momentum $\mathbf{q}=0$, applying the scissor operator estimated in Table~\ref{tab:IND_excitons_values}. In this way we found the excitons already present in the unit cell and new lower 4 times degenerate \emph{dark} excitons, corresponding to the indirect excitons mapped at $\Gamma$. Being $dark$, the indirect excitons do not contribute to the optical response if the coupling with phonons was not explicitly taken into account.
\item We map the phonon-modes at $\qq=K$ in the supercell and generate 12 supercells one for each phonon mode, with atoms displaced along the phonon eigenmodes times 0.1 Bohr. 
\item For each displaced supercell we diagonalized the BSE where single particle energies have been corrected using the previously estimated scissor operator to reproduce the exact exciton energy at momentum $K$ in the primitive cell. Notice that we repeated the calculation by orienting the electric field, which enter into the exciton dipole matrix elements $T_S$, in each of the three directions of space  ($\mathbf{\hat r}$=\{$\hat x$, $\hat y$, $\hat z$\}) in order to subsequently calculate a spatial average of the luminescence signal. Moreover, in the displaced supercells we do not recalculate the dielectric constant but we use the equilibrium one. This approximation allows us to speed up calculation and it has been shown not to produce any visible error in the final luminescence spectrum.\cite{lechifflart2022excitons}  
\item Using the BSE results of the displaced supercells we calculated the second order derivatives of the excitonic dipole matrix elements $\frac{\partial^2 |T_S|^2}{\partial R^2_{\lambda\qq} }$ by finite differences, where S is the exciton index and ${\lambda,\qq}$ the phonon mode, as pointed out in Eq.\,\ref{eq:RS}.\cite{yambopy}.  
\item Finally we applied Eq.~\ref{eq:RS} to get the luminescence spectra at $0$ kBar, and spatially averaged on the three Cartesian directions. In Eq.~\ref{eq:RS} we used an excitonic temperature of 75~K and a lattice one of 55~K in analogy with the measurements on $h$BN.\cite{Paleari2019} Notice that in our case the sum over $\qq$ points reduces to the single point $\qq$=$K$.  
\end{enumerate}
This procedure is explained with a working example on the wiki webpage of the Yambo code.\cite{yambowiki}

%============================
\section{Results}
\label{sec:results}
%===========================
In this section we present results for the different properties of $w$BN as a function of the pressure between 0 and 200~kBar.

\begin{figure}
	\centering
 	\includegraphics[width=0.45\textwidth]{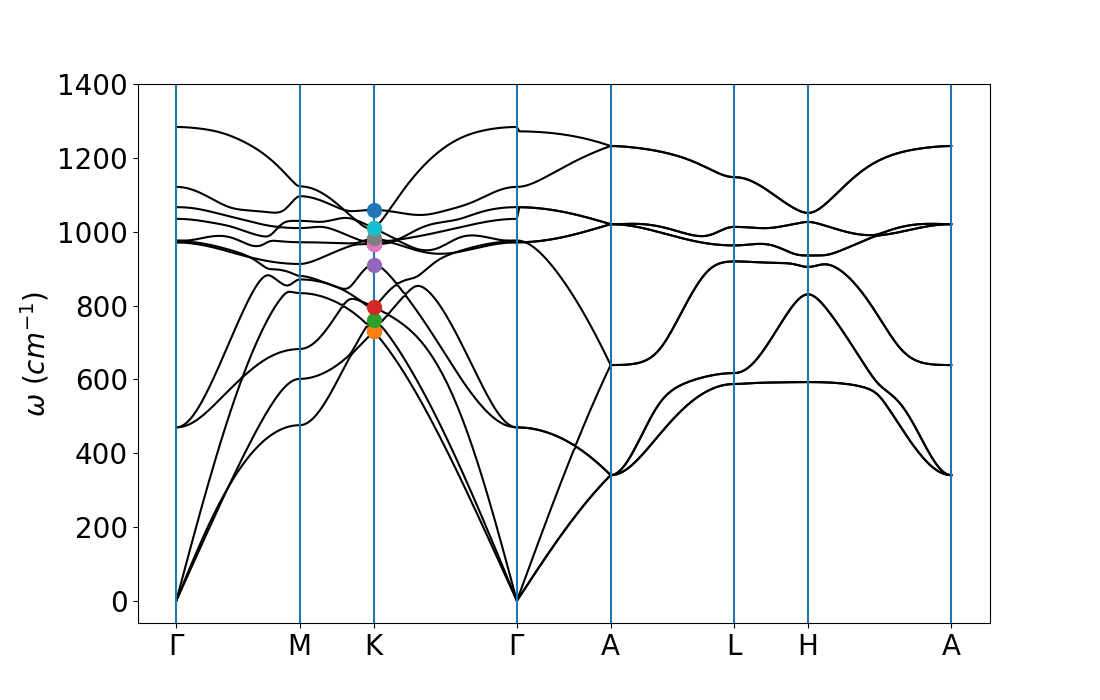}
	\caption{Phonon dispersion of $w$BN at 0 kBar calculated in the framework of DFPT.}
	\label{fig:phonon_disp}
 \end{figure}

\begin{figure}
	\includegraphics[width=0.45\textwidth]{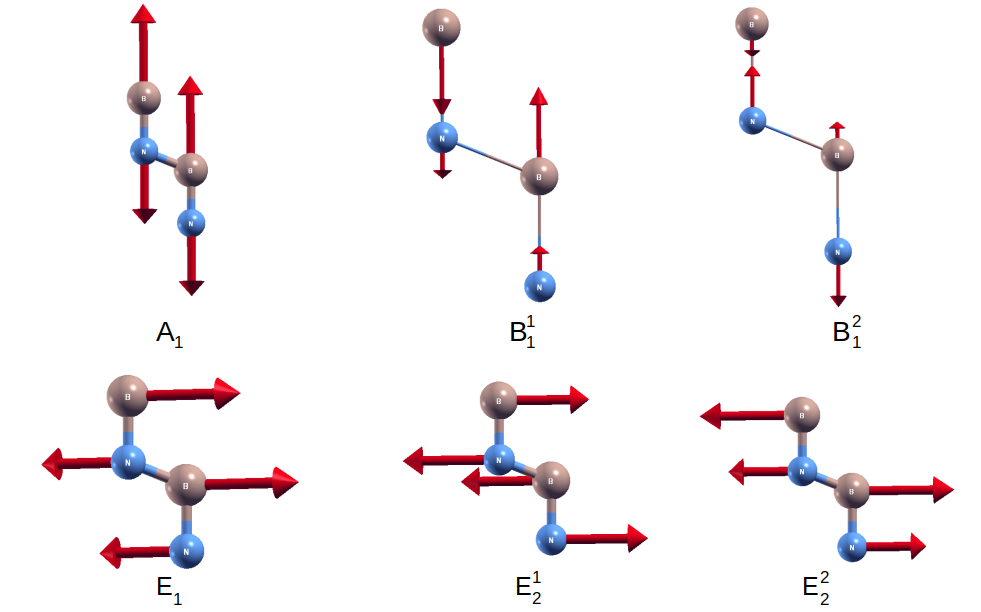}
	\caption{Phonon eigenmodes at $\Gamma$ point and their irreducible representations. The $A_1$ eigenmode is intended to be the same for TO and LO modes, while the here represented $E_1$ mode corresponds to the TO one, the LO being out of plane with respect to the TO mode.}
	\label{fig:phonon modes}
\end{figure}

%=======================
\emph{Vibrational Properties}
%=======================
The phonon dispersion at 0 kBar is represented in Fig.\,\ref{fig:phonon_disp}.  The four atoms in the unit cell give rise to twelve phonon branches for a given \(\boldsymbol{q}\)-point. According to the analysis of the $w$BN symmetry group \(C_{6v}\), phonon modes can be decomposed into $N(A_1 \oplus B_1 \oplus E_1 \oplus E_2)$ modes, where $N=2$. Whose modes $A_1 \oplus E_1$ are acoustic and $A_1 \oplus 2 B_1 \oplus E_1 \oplus 2 E_2$ are optic modes (nine in total). Except for $B_1$ modes which are infrared and Raman forbidden, the others are Raman active modes. Out of them $A_1$ and $E_1$ modes are also infrared active. The $A_1$ and $E_1$ modes are each split into LO and TO components because of the macroscopic electric field associated  with the motion of the longitudinal-optical phonon vibrations and the consequent dependence on the \(\boldsymbol{q}\)-point of the Born effective charges. Phonon modes are identified in Fig.~\ref{fig:phonon modes} according to notation used in Ref.\,\onlinecite{Gorczyca1995}. \\

\begin{figure}
	\includegraphics[width=0.45\textwidth]{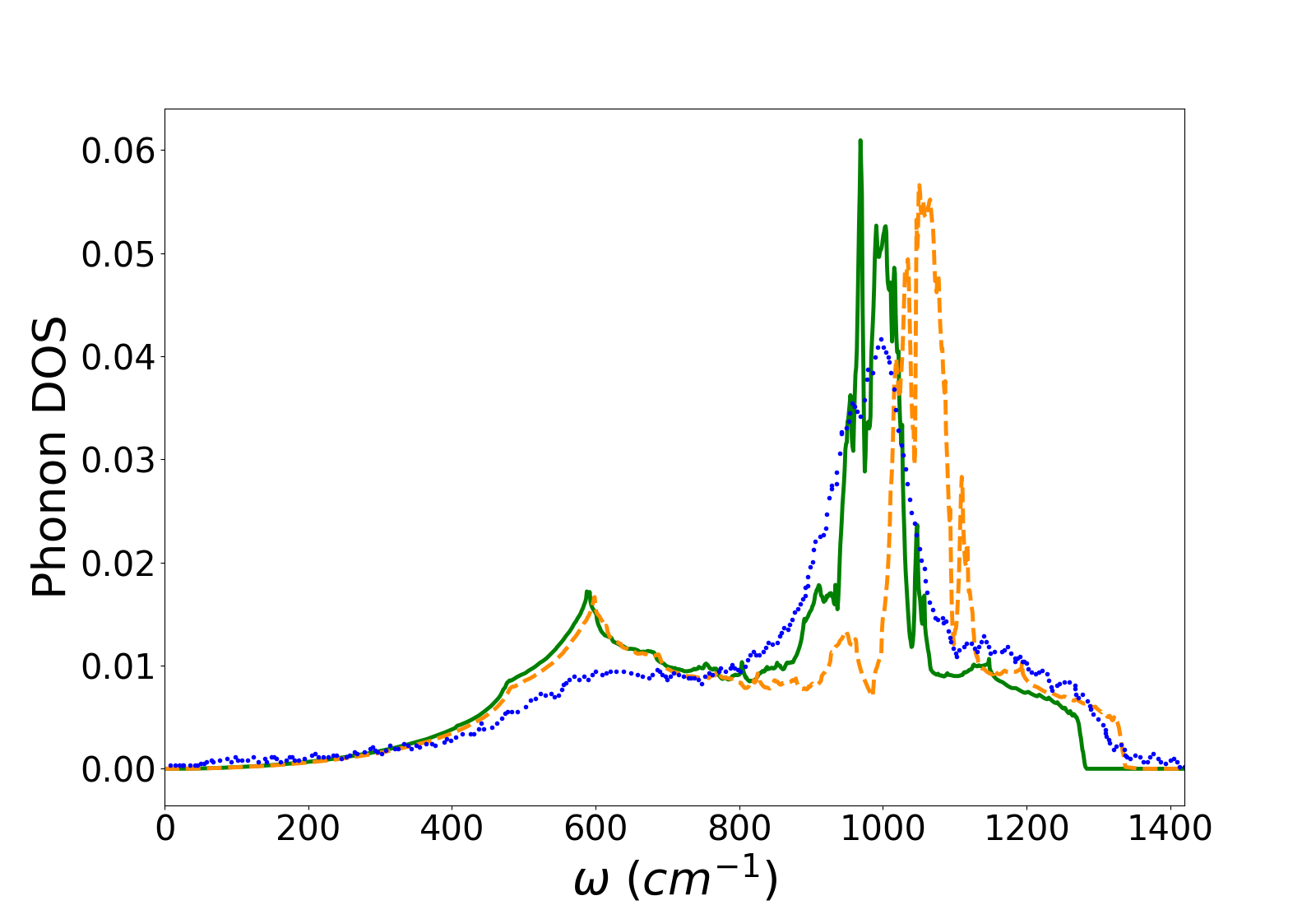}
	\includegraphics[width=0.45\textwidth]{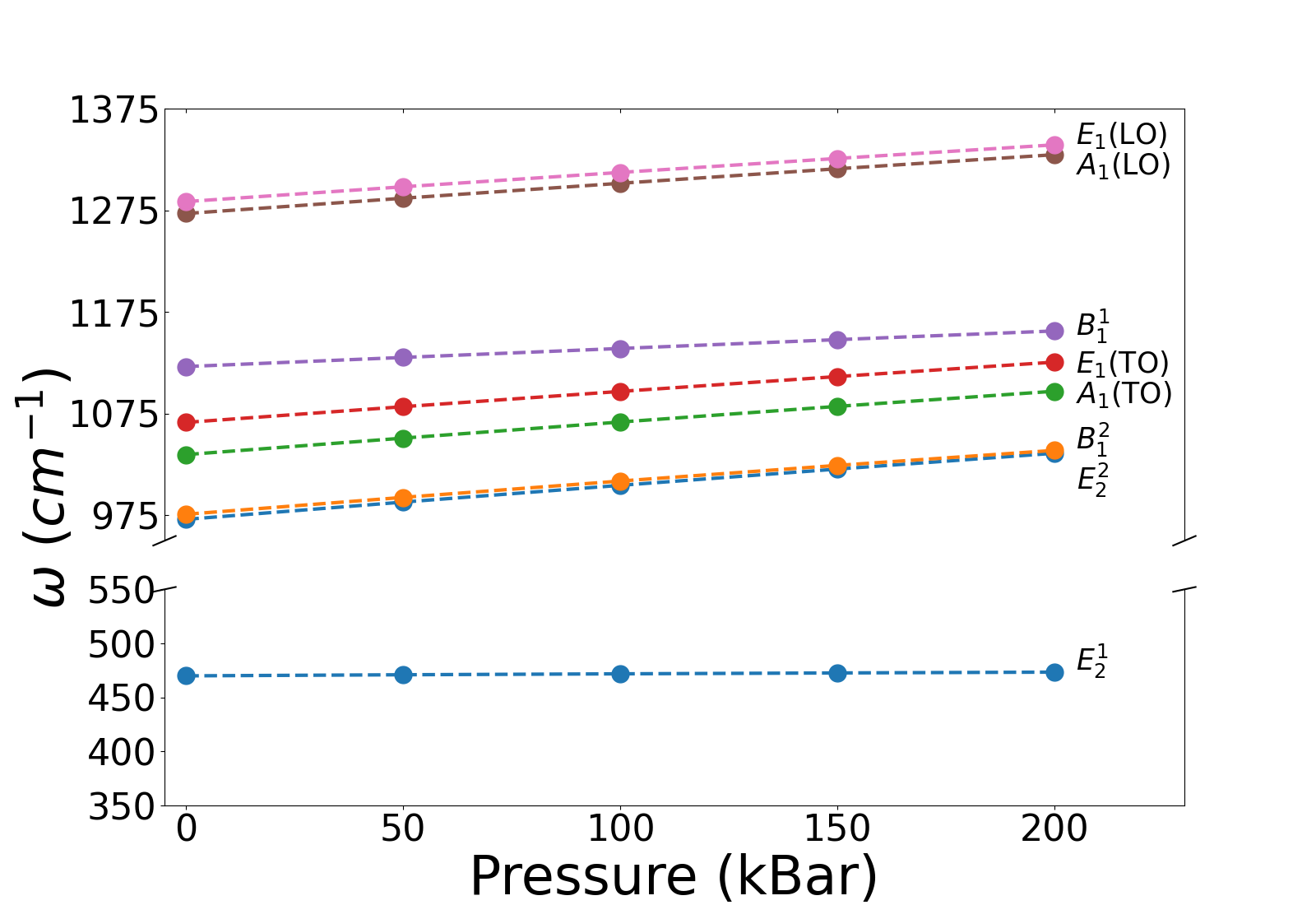}
	\caption{Top panel: Calculated phonon density of states for 0 (green, solid line) and 200 kBar (orange, dashed line) and experimental phonon density of states (blue, dotted line) extracted from inelastic x-ray scattering in Refs. \cite{BOSAK20061661}, multiplied by 2 to make the comparison easier.  Bottom panel: Optical phonon frequencies at \(\Gamma\) point in $w$BN as a function of hydrostatic pressure. The first three acoustic modes are omitted.}
	\label{fig:phdos_phonon_energies_pressures}
\end{figure}

We are going to examine now the evolution of phonon modes as a function of pressure. 
As hydrostatic pressure increases from 0 to 200 kBar, with a consequent compression of the entire crystal structure, phonon frequencies in general increase as well as expected. The general tendency is captured by the phonon density of state represented in Fig.~\ref{fig:phdos_phonon_energies_pressures} (top panel), obtained by sampling the Brillouin zone on a uniform 30$\times$30$\times$20 $\qq$-point grid. We observe that the phonon density of states, calculated at $0$ kBar, agrees with the experimental DOS and that low energy optical phonons (in the frequency range $400-800$ $cm^{-1}$) are slightly affected when the pressure is increased from 0 to 200 kBar, while the high energy optical phonons are significantly shifted.

This point is highlighted in Fig.~\ref{fig:phdos_phonon_energies_pressures} (bottom panel) where we report the phonon energies at the \(\Gamma\) point as a function of pressure (with the exception of the first three acoustic modes). Note that, in contrast to all the other modes which have a similar slope, the double degenerate $E_{2}^{1}$ mode is less sensitive to changes in pressure. This is consistent with the fact that this mode is associated with the phase transition to the zincblende phase ($c$BN). The $E_{2}^{1}$ mode is in fact the result of the folding of the cubic phase transverse acoustic phonon branch onto $\Gamma$ in the wurtzite phase. Interestingly the frequency associated with this mode is compatible with the phonon replica ($\sim$ 60 meV) observed in the cathodoluminescence spectra of $c$BN, supporting then the hypothesis that $w$BN domains are present in $c$BN after phase transformation under high pressure conditions\,\cite{Tararan2018}.  
%This may be due to the fact that in the \(E_{2}^{1}\) mode the projection on the axis of the covalent bondings of the relative atomic displacement of Boron and Nitrogen is smaller compared to the other modes, therefore it's less sensitive to a change of the atomic distances produced by a variation of pressure.\\
In Tab.~\ref{tab:Gph_irrep} we report the calculated phonon frequencies at $\Gamma$ point at $0$ kBar, which are in agreement with previous results\cite{Karch1997,segura2021tuning}.   
\begin{table}
	\centering
	\begin{tabular}{|c|c|c|c|}
		\hline
		Irr. repr \(E_{\Gamma}\) & This work & Ref.~\onlinecite{Segura2019-nonreversiblewbn} & Ref.~\onlinecite{Karch1997} \\
		\hline
		 \(E_{2}^{1}\)  & 469.96 & 475.1 & \\
		 \(E_{2}^{2}\)  & 971.08 & 978.3 & \\
		 \(B_{1}^{1}\)  & 976.17 & 981.3 & \\
		 \(A_{1}(TO)\)  & 1034.7 & 1043.4 & 1006 \\
		 \(E_{1}(TO)\)  & 1066.4 & 1075.9 & 1053 \\
		 \(B_{1}^{2}\)  & 1121.3 & 1131.3 & \\
		 \(A_{1}(LO)\)  & 1271.8 & 1278.9 & 1258 \\
		 \(E_{1}(LO)\)  & 1283.6 & 1291.5 & 1281 \\
		\hline
	\end{tabular}
	\caption{Longitudinal and transverse zone-center phonon frequencies of $w$BN at 0 kBar in $cm^{-1}$, compared with Refs.~\onlinecite{Karch1997,Segura2019-nonreversiblewbn}. The $A_1$ mode LO-TO splitting is obtained by sampling the Brillouin zone along the $\Gamma$-A and $\Gamma$-K directions respectively.} 
	\label{tab:Gph_irrep}
\end{table}
Here we observed that the $E_1$ LO-TO splitting amounts to $217.2$ $cm^{-1}$ at 0 kBar while it decreases up to $213.7$ $cm^{-1}$ at 200 kBar (see SM\,\cite{SM}, Table S2), accordingly to the decrease of the Born effective charges and the dielectric tensor matrix elements (see SM\cite{SM}, Table S3). This finding goes in the same direction as what happens for wurtzite phase GaN and AlN\cite{Perlin1999,Gogni2001,Gorczyca1995} and other III-V compounds\cite{Reparaz2018}. As pressure increases such splitting is always more pronounced than the $E_1$-$A_1$ splitting (which amounts to $11.7\,\text{cm}^{-1}$ at 0 kBar and $9.7\,\text{cm}^{-1}$ at 200 kBar, see SM\cite{SM}, Table S2). Therefore $w$BN belong to crystals of class I, according to Refs.\,\onlinecite{loudon1964,Feldman1968}, in which the electrostatic Coulomb forces dominate over the anisotropy of the short-range interatomic forces.

%======================================
\emph{Electronic band structure and quasi-particle properties}
%======================================

\begin{figure}[t]
	\centering
	\includegraphics[scale=0.33]{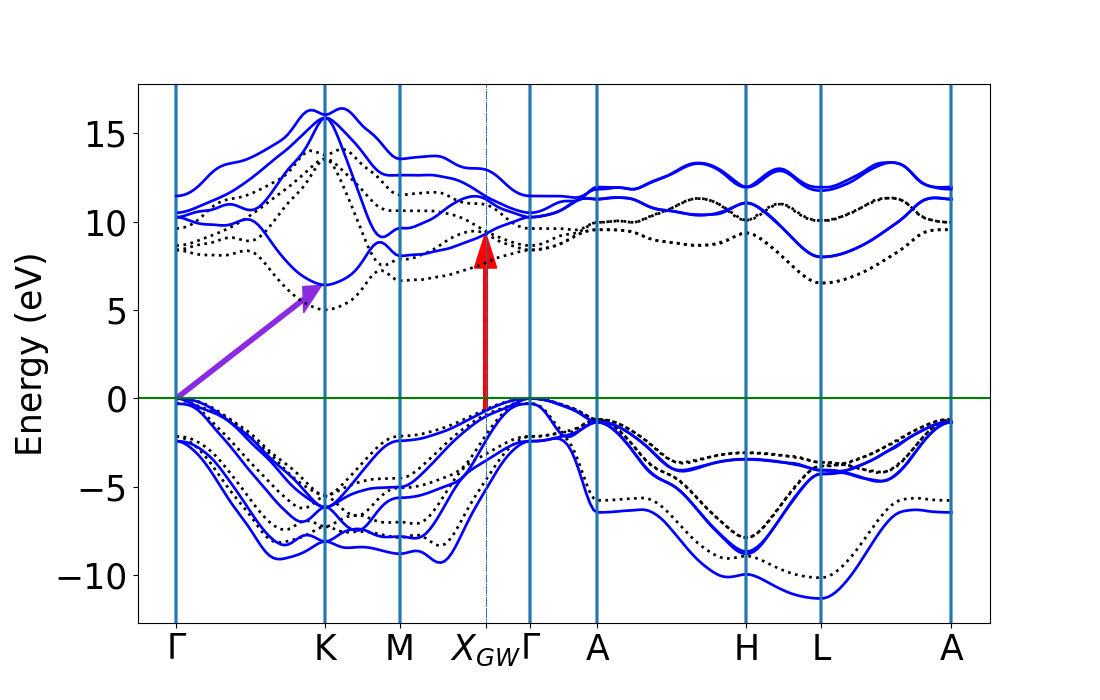}
	\caption{DFT (black points) and QP-corrected electronic (solid blue line) band structure of $w$BN at 0 kBar. Arrows highlight the transitions where direct (red) and indirect (violet) gaps occur at QP level. Only four bands above and below the band gap are shown.}
	\label{fig:bandstructQP}
\end{figure}

\begin{figure}[t]
	\centering
	\includegraphics[width=0.53\textwidth]{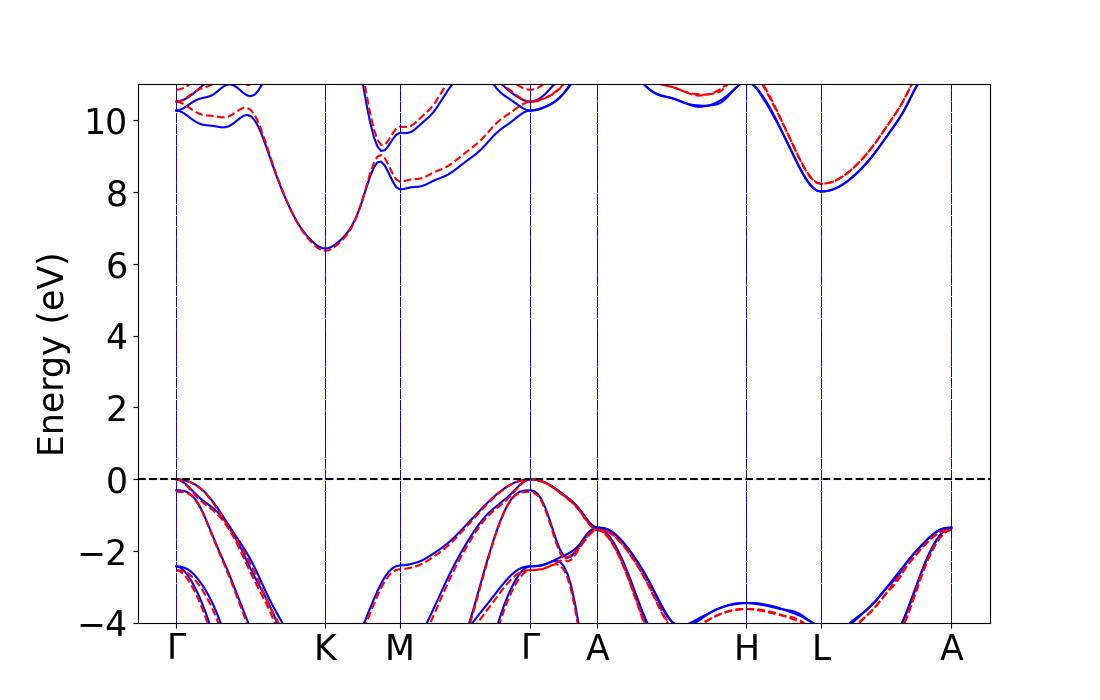}
	\caption{QP corrected band structure at 0 kBar (blue solid line) and 200 kBar (red dashed line).}
	\label{fig:BS_QP_vs_P}
\end{figure}

Here we focus on the evolution of electronic properties as a function of pressure. At $0$ kBar we found that at the DFT level the lowest energy gap occurs for a transition between $\Gamma$ and K points of the Brillouin zone, and amounts to $5.01$ eV. The direct gap of $8.3$ eV, in agreement with previous results\cite{Christensen1994}, occurs at a {\bf k}-point close to $\Gamma$ that we will call hereafter $X_{DFT}=(0,\nicefrac{1}{12},0)$.  
The quasiparticle (QP) correction in GW approximation widens the indirect and direct gap of $1.4$ and $1.72$ eV respectively, at 0 kBar. The {\bf k}-point where the direct transition occurs (direct gap transition) shifts a little bit further away from the $\Gamma$ point to the new {\bf k}-point of coordinates $X_{GW}=(0,\nicefrac{1}{6},0)$. The DFT and QP band structures together with direct and indirect band gap transitions are highlighted in Fig.\,\ref{fig:bandstructQP}. While in Fig.\,\ref{fig:BS_QP_vs_P} the QP corrected band structure at the extreme values of the considered pressure range are shown. While the first valence bands are essentially unaffected, along $\Gamma$-M line (where the direct transition occurs) the lowest conduction band shifts up, while at K it is slightly shifted down. Such an observation opens the discussion on the evolution of the band gaps with the applied hydrostatic pressure.      

\begin{figure}
	\centering
	\begin{minipage}[b]{0.46\columnwidth}
         \centering
         \includegraphics[width=\textwidth]{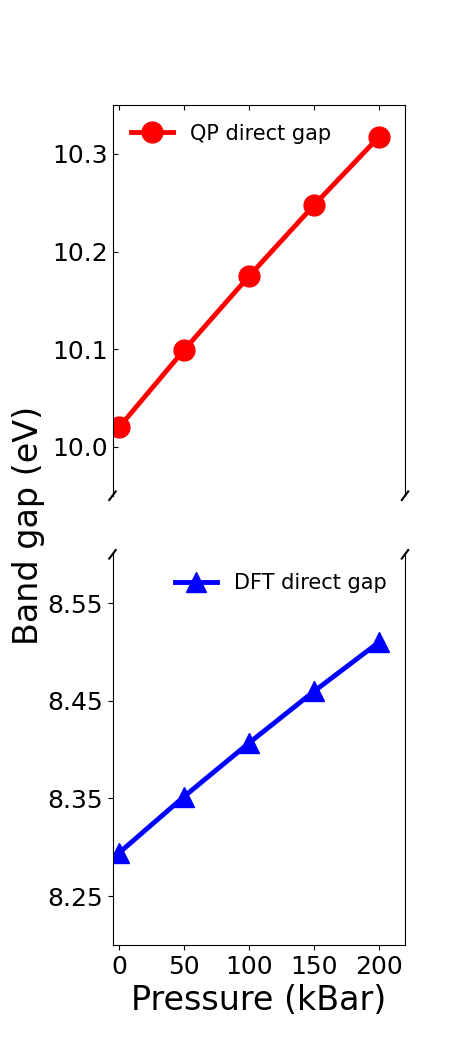}
     \end{minipage}
     \hfill
     \begin{minipage}[b]{0.46\columnwidth}
         \centering
         \includegraphics[width=\textwidth]{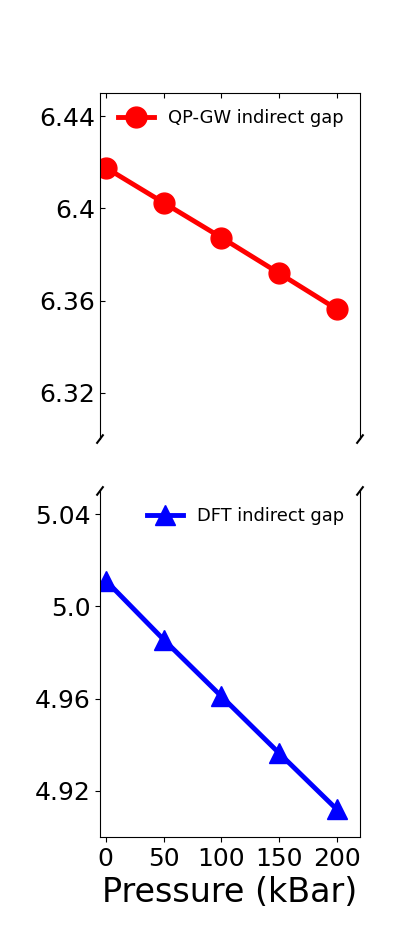}
     \end{minipage}
     \caption{DFT and QP-corrected direct and indirect gaps as a function of pressure. }
     \label{fig:bandgaps}
\end{figure}

As pressure increases the {\bf k}-space positions ($X_{DFT}$ and $X_{GW}$) of the direct gap do not change and in the 0-200 kBar range the QP correction itself is almost constant. On the other hand direct and indirect band gaps undergo a different behavior with pressure (see Fig.\,\ref{fig:bandgaps}). We observe that while the first tends to increase the latter does the opposite. This is in contrast of what observed for direct and indirect band gaps in $h$BN under pressure.\cite{segura2021tuning}  
We ascribed this behaviour to the different character of the band edges, that we deduced by projecting the Kohn-Sham states on selected atomic orbitals of B, N atoms (see SM\cite{SM}, Fig.\,S3).
The states taken into consideration are $2s$ and $2p$ for each of the four atoms in the primitive cell. 
%In Table~\ref{gaps at different pressures DFT} and \ref{gaps at different pressures QP} the behaviour of both direct and indirect gap are summarized.
%\begin{table}[h]
%	\begin{tabular}[c]{c|c|c}
%		\centering
%		       Pressure & Ind. Gap &  Dir Gap (X point)  \\ 
%		         (kBar) &  (eV)    &   (eV)  \\ \hline\hline
%		         0      &  5.011   &   8.29\\
%		         50     &  4.986   &   8.35\\
%		        100     &  4.961   &   8.41\\
%		        150     &  4.936   &   8.46\\
%		        200     &  4.912   &   8.51\\
%	\end{tabular}                
%	    \label{tab:gap_vs_P}
%	    \caption{Overview over direct and indirect DFT band gaps values in wurtzite-BN.}
%\end{table}
The electrons contributing to conduction bands are mostly from the B atoms while the electrons contributing to the valence ones come from the N atoms, this is in particular true for {\bf k}-points very close $\Gamma$ points. However at the K point the contribution comes from a mixture of the orbitals of two atoms, making the behaviour of the indirect gap different from the direct one. 

\begin{figure}
	\centering
	\includegraphics[width=0.5\textwidth]{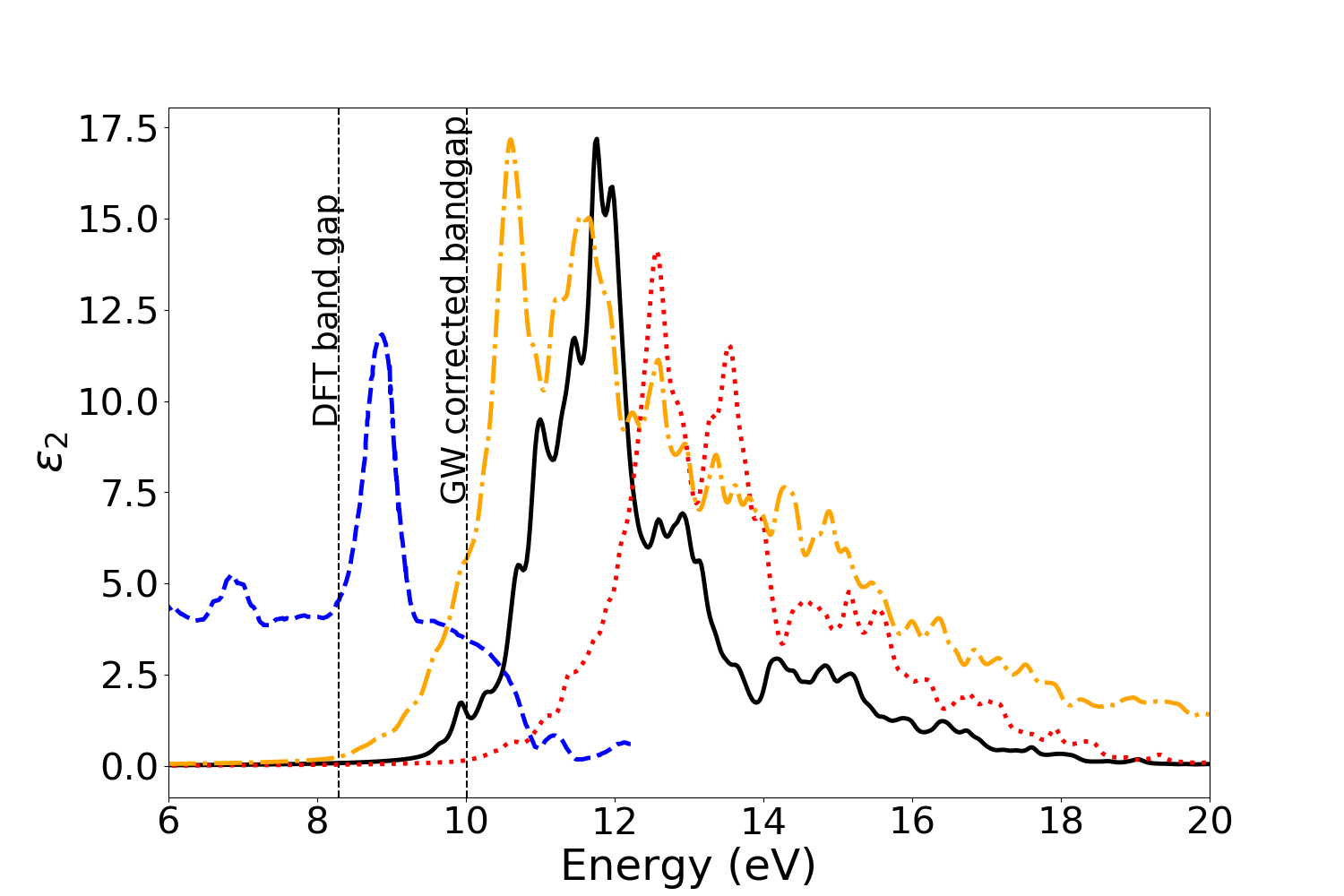}
	\caption{Quasi-particle (solid black), DFT based independent particle (dashed-dotted orange), GW based independent particle (dotted red) spectra of the imaginary part of \(\epsilon_{M}\) at 0 kBar. Dashed vertical lines mark the position of the GW-corrected and the DFT band gaps in both main figure and inset. Experimental data (dashed blue) are extracted from Ref.\,\onlinecite{Yixi1994}}.
	\label{fig:Im_epsilon_ip_vs_qp}
\end{figure}

\emph{Light absorption}
%======================================
In Fig.\,\ref{fig:Im_epsilon_ip_vs_qp} we plot a comparison between the imaginary part of the dielectric function calculated at the independent particle (IP) level, both from DFT and GW band structure, and by including the electron-hole interaction. Such interaction modifies the absorption signal, resulting in both a redistribution of the main features intensity and the appearance of an excitonic peak below the GW-corrected band gap at 9.93 eV, absent at the IP level.    

Our analysis of the excitonic structure of the absorption spectrum revealed the presence of a low energy \emph{dark} exciton and two degenerate excitons at 9.59 eV with a small dipole. The first bright exciton, which determines the optical gap at 0 kBar, is also doubly degenerate and has an energy of 9.64 eV, a value quite close to that one calculated for $c$BN\,\cite{Tararan2018}. In Fig.~\ref{fig:Im_epsilon_ip_vs_qp} we also plot the experimental dielectric constant measured in Ref.~\onlinecite{Yixi1994}. The authors affirm that the optical gap was found to be at $8.7\pm0.5$ eV. Within the error bar such a value could correspond to our first exciton, which has a non-negligible dipole at 9.64 eV. Overall, the comparison is rather delicate, the experimental spectrum does not extend to higher energies. The discrepancy in energy between our theoretical predictions and experimental results can only partially be attributed to the absence of electron-phonon interaction in our theoretical approach.\cite{kawai2014electron} However it is also possible that the distinct features observed in the experimental spectrum are caused by the presence of defects, which cannot be ruled out at this time. In this case, further studies are necessary to determine the source of these signals, and luminescence measurements could easily discriminate between defect and bulk states due to the indirect nature of this material, see next section.

\begin{figure}
	\centering
	\includegraphics[scale=0.3]{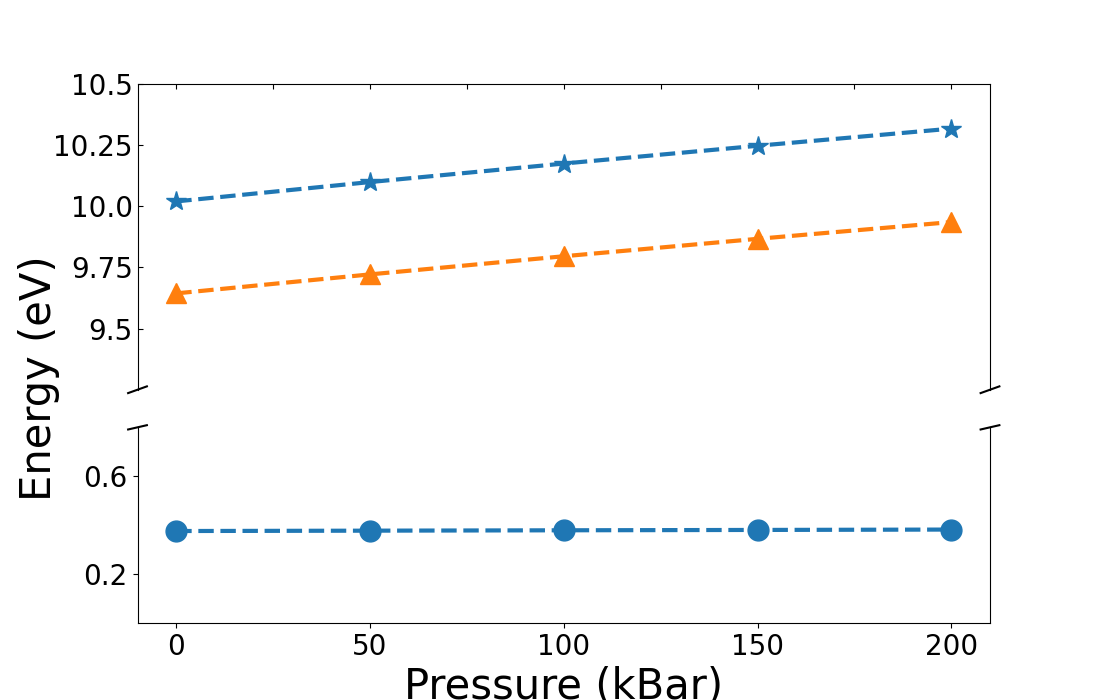}
	\caption{Quasiparticle-corrected direct band gap (blue stars line), first bright exciton level (orange triangles line) and the binding energy (blue dots line) as a function of pressure. }
	\label{fig:binding_energy_exc_gamma}
\end{figure}

We then proceed to analyse the effect of pressure on the lowest excitonic peaks. As shown in Fig.\,\ref{fig:binding_energy_exc_gamma}, optical excitation shift linearly towards high energies from 0 kBar to 200 kBar, closely following the linear behaviour of the quasi-particle corrected direct band gap. Consequently, the exciton binding energy $(E_{bind}=E_{gap}-E_{exc})$ remains almost constant with pressure, with a value of 0.38 eV at 0 kBar. This behaviour can be explained by considering how screening affects the electron-hole pair. In fact excitons in $w$BN are highly delocalised well beyond the boundaries of the unit cell. Thus, changing the geometry of the system has less effect on the screening which affects their electron-hole interaction.

%======================================
\emph{Phonon assisted luminescence}
%======================================
\begin{figure}
   \centering
	\includegraphics[scale=0.3]{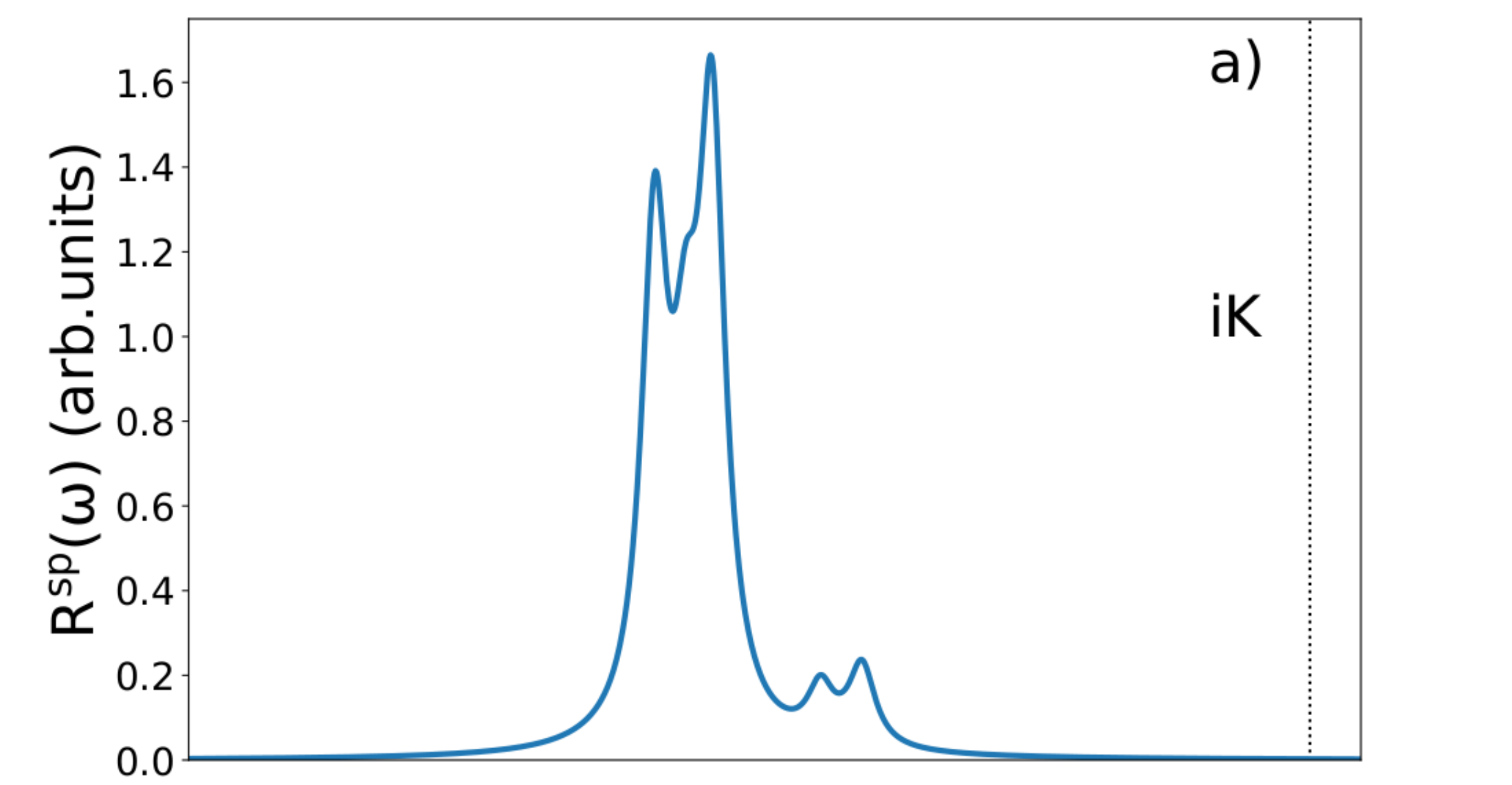} 
	\includegraphics[scale=0.3]{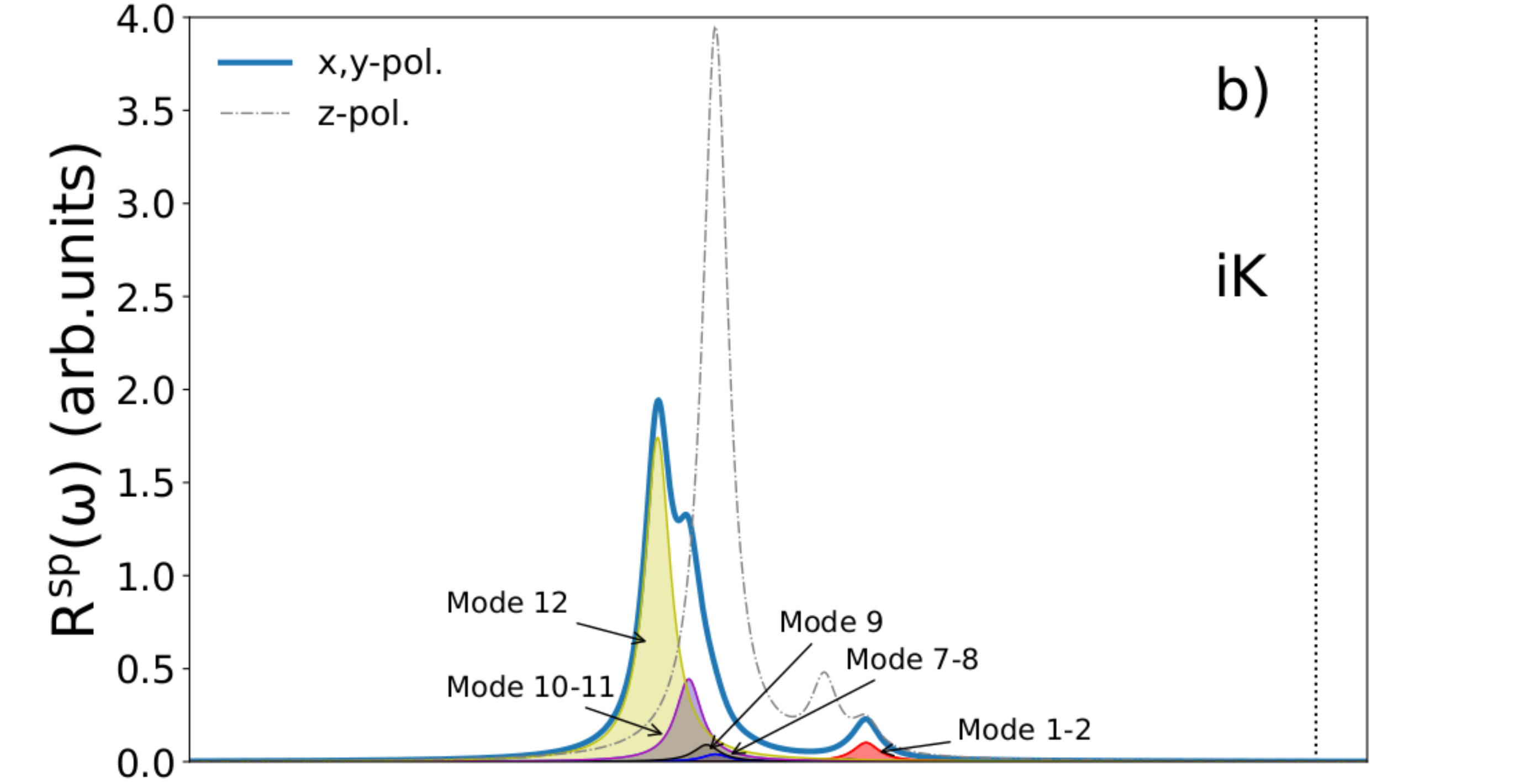}
	\includegraphics[scale=0.3]{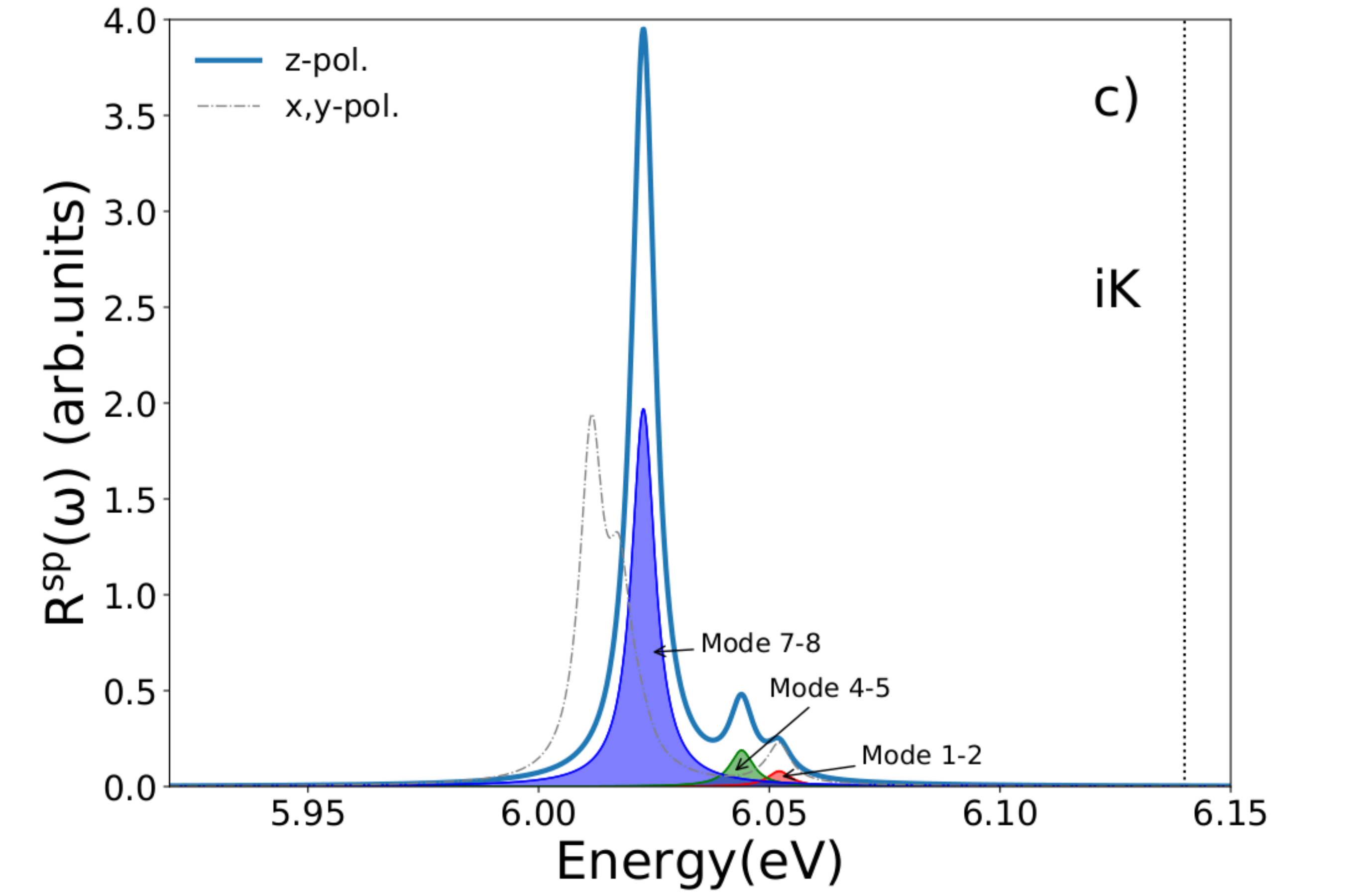}
        \caption{a) Phonon-assisted luminescence at $0$ kBar, when the luminescence has been averaged across the three Cartesian directions. The dotted vertical line indicates the energy of the indirect exciton (iK). In panel b) and c) we separated the contributions of the basal plane Cartesian coordinates (average over the x and y directions) and the z-axis. In both cases the contributions of single phonon modes have been specified.  Note that a couple of phonon modes (the 3rd and the 6th) have a much lower contribution compared to the others and are not shown in the figure.}
	\label{fig:PHass_Lumin}
\end{figure}
In Fig.~\ref{fig:PHass_Lumin} we report phonon-assisted luminescence calculated using the theoretical methods presented in Sec.~\ref{sec:theo} and ~\ref{sec:comp}.\\
The luminescence is calculated at $0$ kBar, and spatially averaged on the three Cartesian directions, panel $(a)$ in  Fig.\,\ref{fig:PHass_Lumin}\,. One observes four main peaks, at $6.012$, $6.022$ (plus a shoulder at $6.018$ eV), $6.044$ and $6.053$ eV. We first decomposed the spectrum into the contribution coming from the basal plane components and that from the z-axis as shown in panels $(b)$ and $(c)$ of Fig.\,\ref{fig:PHass_Lumin}\,. This allows us to interpret the origin of the four main peaks. Additionally the contribution of each phonon mode at $K$ point of the Brillouin Zone is highlighted, allowing to identify each peak as the result of the coupling of the lowest indirect exciton $iK$ and a precise phonon mode at the point $K$ of the BZ. We found that for light polarized in the basal plane the indirect exciton couples mostly with the highest-energy optical modes and to the lowest energy phonon modes to give origin to the peak at $6.012$ eV, the shoulder at $6.018$ eV and the peak at $6.053$ eV. On the other hand when light is polarized along the $z$-axis, the coupling occurs with the $7^{th}$-$8^{th}$, the $4^{th}$-$5^{th}$ and slightly with the $1^{st}$-$2^{nd}$ phonon modes, and originate to the intense signal at $6.022$ eV, to that at $6.044$ and $6.053$ eV respectively. \\ 
On the basis of the evolution as a function of pressure of the QP direct band gap and of the direct exciton energy (see Figs.\,\ref{fig:bandstructQP}-\ref{fig:binding_energy_exc_gamma}), we deduce that the indirect exciton energy will decrease as pressure increases. Beside that, the phonon frequencies at $K$ point (see SM\cite{SM}, Fig.\,S2 and Table S4) tend to increase with pressure even if with a relatively different rates when optic and acoustic modes are compared. Putting all this information together we expect a slight red shift with pressure of the luminescence spectrum accompanied by a small deviation of the two main peaks, since the originally located peak at $6.02$ eV will red shift more than the other at $6.05$ eV, because of the contribution coming from different phonon modes. 
As mentioned above, the remarkable light-emitting properties of $h$BN have been the subject of extensive research in recent years. Recent studies have shown that its high light-emitting efficiency is due to its flat band nature at the band edges\,\cite{elias2021flat}. Although experimental luminescence data are not currently available, it is possible to speculate that the light-emitting properties of $w$BN would be more similar to those of diamond, given its more parabolic band nature\,\cite{schue2019bright}. As a result, the inclusion of $w$BN in $h$BN may reduce its luminescence efficiency. We expect that this prediction could be verified in future work.

\section{Conclusions}
%=================================
This study presents an in-depth analysis of the properties of $w$BN over a wide range of pressures. Our findings reveal that $w$BN is a large indirect gap material, with a predicted direct gap near the $\Gamma$ point, which increases with pressure, while the indirect gap between $\Gamma$$\rightarrow$K decreases. We also explored the vibrational properties of $w$BN and have discovered that the frequency associated with the $E^1_2$ mode aligns with the phonon replica observed in the cathodoluminescence spectra of $c$BN. In terms of optical properties, our results show that the first optically active excitations occur at 9.64~eV, below the GW gap.
%, quite close to that of $c$BN. A more intense excitonic bound state is visible below the GW gap at about 9.93~eV. 
It is important to consider the electron-hole interaction to interpret the experimental measurements, which display peaks below the GW gap. At the IP level the onset of the light absorption is high in energy, and not compatible with the experimental observations. 
We also studied light emission in $w$BN, and found that phonon-mediated emission is expected to occur at around 6~eV, but it will be less intense than in $h$BN due to the non-flat bands of $w$BN.
%Due to the large difference between the direct and indirect gap, we expect phonon mediated emission to occur at around 6~eV and it will be less intense than in hBN due to the non-flat bands of $w$BN. 
Overall, this study, along with new techniques for the synthesis of $w$BN, can aid in the identification of $w$BN through luminescence measurements and promote its use in optoelectronics in the future.  

%======================
\begin{acknowledgments}
%==============
EC acknowledges the support of the French Agence Nationale de la Recherche (ANR) under reference ANR-20-CE47-0009-01-NOTISPERF. Centre de Calcul Intensif d'Aix-Marseille is acknoledged for granting access to its high performance computing resources. This work was granted access to the HPC/AI resources of TGCC under the allocation 2022-AD010913493 made by GENCI. The research leading to these results has  received funding from the European Union Seventh Framework Program under grant agreement no. 785219 Graphene Core2. This publication is based upon work from COST Action TUMIEE CA17126, supported by COST (European Cooperation in Science and Technology). The authors acknowledge A. Saul and K. Boukari for the management of the computer cluster \emph{Rosa}.
\end{acknowledgments}

\bibliographystyle{apsrev4-1}
\bibliography{wBN}% Produces the bibliography via BibTeX.

\begin{thebibliography}{54}%
\makeatletter
\providecommand \@ifxundefined [1]{%
 \@ifx{#1\undefined}
}%
\providecommand \@ifnum [1]{%
 \ifnum #1\expandafter \@firstoftwo
 \else \expandafter \@secondoftwo
 \fi
}%
\providecommand \@ifx [1]{%
 \ifx #1\expandafter \@firstoftwo
 \else \expandafter \@secondoftwo
 \fi
}%
\providecommand \natexlab [1]{#1}%
\providecommand \enquote  [1]{``#1''}%
\providecommand \bibnamefont  [1]{#1}%
\providecommand \bibfnamefont [1]{#1}%
\providecommand \citenamefont [1]{#1}%
\providecommand \href@noop [0]{\@secondoftwo}%
\providecommand \href [0]{\begingroup \@sanitize@url \@href}%
\providecommand \@href[1]{\@@startlink{#1}\@@href}%
\providecommand \@@href[1]{\endgroup#1\@@endlink}%
\providecommand \@sanitize@url [0]{\catcode `\\12\catcode `\$12\catcode
  `\&12\catcode `\#12\catcode `\^12\catcode `\_12\catcode `\%12\relax}%
\providecommand \@@startlink[1]{}%
\providecommand \@@endlink[0]{}%
\providecommand \url  [0]{\begingroup\@sanitize@url \@url }%
\providecommand \@url [1]{\endgroup\@href {#1}{\urlprefix }}%
\providecommand \urlprefix  [0]{URL }%
\providecommand \Eprint [0]{\href }%
\providecommand \doibase [0]{http://dx.doi.org/}%
\providecommand \selectlanguage [0]{\@gobble}%
\providecommand \bibinfo  [0]{\@secondoftwo}%
\providecommand \bibfield  [0]{\@secondoftwo}%
\providecommand \translation [1]{[#1]}%
\providecommand \BibitemOpen [0]{}%
\providecommand \bibitemStop [0]{}%
\providecommand \bibitemNoStop [0]{.\EOS\space}%
\providecommand \EOS [0]{\spacefactor3000\relax}%
\providecommand \BibitemShut  [1]{\csname bibitem#1\endcsname}%
\let\auto@bib@innerbib\@empty
%</preamble>
\bibitem [{\citenamefont {Mao}\ \emph {et~al.}(2018)\citenamefont {Mao},
  \citenamefont {Chen}, \citenamefont {Ding}, \citenamefont {Li},\ and\
  \citenamefont {Wang}}]{RevModPhys.90.015007}%
  \BibitemOpen
  \bibfield  {author} {\bibinfo {author} {\bibfnamefont {H.-K.}\ \bibnamefont
  {Mao}}, \bibinfo {author} {\bibfnamefont {X.-J.}\ \bibnamefont {Chen}},
  \bibinfo {author} {\bibfnamefont {Y.}~\bibnamefont {Ding}}, \bibinfo {author}
  {\bibfnamefont {B.}~\bibnamefont {Li}}, \ and\ \bibinfo {author}
  {\bibfnamefont {L.}~\bibnamefont {Wang}},\ }\href {\doibase
  10.1103/RevModPhys.90.015007} {\bibfield  {journal} {\bibinfo  {journal}
  {Rev. Mod. Phys.}\ }\textbf {\bibinfo {volume} {90}},\ \bibinfo {pages}
  {015007} (\bibinfo {year} {2018})}\BibitemShut {NoStop}%
\bibitem [{\citenamefont {Segura}\ \emph {et~al.}(2021)\citenamefont {Segura},
  \citenamefont {Cusc{\'o}}, \citenamefont {Attaccalite}, \citenamefont
  {Taniguchi}, \citenamefont {Watanabe},\ and\ \citenamefont
  {Artús}}]{segura2021tuning}%
  \BibitemOpen
  \bibfield  {author} {\bibinfo {author} {\bibfnamefont {A.}~\bibnamefont
  {Segura}}, \bibinfo {author} {\bibfnamefont {R.}~\bibnamefont {Cusc{\'o}}},
  \bibinfo {author} {\bibfnamefont {C.}~\bibnamefont {Attaccalite}}, \bibinfo
  {author} {\bibfnamefont {T.}~\bibnamefont {Taniguchi}}, \bibinfo {author}
  {\bibfnamefont {K.}~\bibnamefont {Watanabe}}, \ and\ \bibinfo {author}
  {\bibfnamefont {L.}~\bibnamefont {Artús}},\ }\href {\doibase
  10.1021/acs.jpcc.1c02082} {\bibfield  {journal} {\bibinfo  {journal} {The
  Journal of Physical Chemistry C}\ }\textbf {\bibinfo {volume} {125}},\
  \bibinfo {pages} {12880} (\bibinfo {year} {2021})}\BibitemShut {NoStop}%
\bibitem [{\citenamefont {Akasaki}\ \emph {et~al.}(2014)\citenamefont
  {Akasaki}, \citenamefont {Amano}, \citenamefont {Nakamura},\ and\
  \citenamefont {Nakamura}}]{akasaki2014nobel}%
  \BibitemOpen
  \bibfield  {author} {\bibinfo {author} {\bibfnamefont {I.}~\bibnamefont
  {Akasaki}}, \bibinfo {author} {\bibfnamefont {H.}~\bibnamefont {Amano}},
  \bibinfo {author} {\bibfnamefont {S.}~\bibnamefont {Nakamura}}, \ and\
  \bibinfo {author} {\bibfnamefont {S.}~\bibnamefont {Nakamura}},\ }\href
  {https://www.kva.se/app/uploads/2014/10/globalassets-priser-nobel-2014-fysik-scibackfyen14.pdf}
  {\bibfield  {journal} {\bibinfo  {journal} {The Royal Swedish Academy of
  Science}\ } (\bibinfo {year} {2014})}\BibitemShut {NoStop}%
\bibitem [{\citenamefont {Watanabe}\ \emph {et~al.}(2004)\citenamefont
  {Watanabe}, \citenamefont {Taniguchi},\ and\ \citenamefont
  {Kanda}}]{Watanabe2004}%
  \BibitemOpen
  \bibfield  {author} {\bibinfo {author} {\bibfnamefont {K.}~\bibnamefont
  {Watanabe}}, \bibinfo {author} {\bibfnamefont {T.}~\bibnamefont {Taniguchi}},
  \ and\ \bibinfo {author} {\bibfnamefont {H.}~\bibnamefont {Kanda}},\ }\href
  {\doibase 10.1038/nmat1134} {\bibfield  {journal} {\bibinfo  {journal}
  {Nature Materials}\ }\textbf {\bibinfo {volume} {3}},\ \bibinfo {pages} {404}
  (\bibinfo {year} {2004})}\BibitemShut {NoStop}%
\bibitem [{\citenamefont {Cassabois}\ \emph {et~al.}(2016)\citenamefont
  {Cassabois}, \citenamefont {Valvin},\ and\ \citenamefont
  {Gil}}]{Cassabois2016}%
  \BibitemOpen
  \bibfield  {author} {\bibinfo {author} {\bibfnamefont {G.}~\bibnamefont
  {Cassabois}}, \bibinfo {author} {\bibfnamefont {P.}~\bibnamefont {Valvin}}, \
  and\ \bibinfo {author} {\bibfnamefont {B.}~\bibnamefont {Gil}},\ }\href
  {\doibase 10.1038/nphoton.2015.277} {\bibfield  {journal} {\bibinfo
  {journal} {Nature Photonics}\ }\textbf {\bibinfo {volume} {10}},\ \bibinfo
  {pages} {262} (\bibinfo {year} {2016})}\BibitemShut {NoStop}%
\bibitem [{\citenamefont {Schué}\ \emph {et~al.}(2016)\citenamefont {Schué},
  \citenamefont {Berini}, \citenamefont {Betz}, \citenamefont {Plaçais},
  \citenamefont {Ducastelle}, \citenamefont {Barjon},\ and\ \citenamefont
  {Loiseau}}]{Schue2016}%
  \BibitemOpen
  \bibfield  {author} {\bibinfo {author} {\bibfnamefont {L.}~\bibnamefont
  {Schué}}, \bibinfo {author} {\bibfnamefont {B.}~\bibnamefont {Berini}},
  \bibinfo {author} {\bibfnamefont {A.~C.}\ \bibnamefont {Betz}}, \bibinfo
  {author} {\bibfnamefont {B.}~\bibnamefont {Plaçais}}, \bibinfo {author}
  {\bibfnamefont {F.}~\bibnamefont {Ducastelle}}, \bibinfo {author}
  {\bibfnamefont {J.}~\bibnamefont {Barjon}}, \ and\ \bibinfo {author}
  {\bibfnamefont {A.}~\bibnamefont {Loiseau}},\ }\href {\doibase
  10.1039/C6NR01253A} {\bibfield  {journal} {\bibinfo  {journal} {Nanoscale}\
  }\textbf {\bibinfo {volume} {8}},\ \bibinfo {pages} {6986} (\bibinfo {year}
  {2016})}\BibitemShut {NoStop}%
\bibitem [{\citenamefont {Fossard}\ \emph {et~al.}(2017)\citenamefont
  {Fossard}, \citenamefont {Sponza}, \citenamefont {Schu\'e}, \citenamefont
  {Attaccalite}, \citenamefont {Ducastelle}, \citenamefont {Barjon},\ and\
  \citenamefont {Loiseau}}]{Fossard2017}%
  \BibitemOpen
  \bibfield  {author} {\bibinfo {author} {\bibfnamefont {F.}~\bibnamefont
  {Fossard}}, \bibinfo {author} {\bibfnamefont {L.}~\bibnamefont {Sponza}},
  \bibinfo {author} {\bibfnamefont {L.}~\bibnamefont {Schu\'e}}, \bibinfo
  {author} {\bibfnamefont {C.}~\bibnamefont {Attaccalite}}, \bibinfo {author}
  {\bibfnamefont {F.}~\bibnamefont {Ducastelle}}, \bibinfo {author}
  {\bibfnamefont {J.}~\bibnamefont {Barjon}}, \ and\ \bibinfo {author}
  {\bibfnamefont {A.}~\bibnamefont {Loiseau}},\ }\href {\doibase
  10.1103/PhysRevB.96.115304} {\bibfield  {journal} {\bibinfo  {journal} {Phys.
  Rev. B}\ }\textbf {\bibinfo {volume} {96}},\ \bibinfo {pages} {115304}
  (\bibinfo {year} {2017})}\BibitemShut {NoStop}%
\bibitem [{\citenamefont {Vuong}\ \emph {et~al.}(2018)\citenamefont {Vuong},
  \citenamefont {Liu}, \citenamefont {Van~der Lee}, \citenamefont {Cuscó},
  \citenamefont {Artús}, \citenamefont {Michel}, \citenamefont {Valvin},
  \citenamefont {Edgar}, \citenamefont {Cassabois},\ and\ \citenamefont
  {Gil}}]{vuong2018}%
  \BibitemOpen
  \bibfield  {author} {\bibinfo {author} {\bibfnamefont {T.~Q.~P.}\
  \bibnamefont {Vuong}}, \bibinfo {author} {\bibfnamefont {S.}~\bibnamefont
  {Liu}}, \bibinfo {author} {\bibfnamefont {A.}~\bibnamefont {Van~der Lee}},
  \bibinfo {author} {\bibfnamefont {R.}~\bibnamefont {Cuscó}}, \bibinfo
  {author} {\bibfnamefont {L.}~\bibnamefont {Artús}}, \bibinfo {author}
  {\bibfnamefont {T.}~\bibnamefont {Michel}}, \bibinfo {author} {\bibfnamefont
  {P.}~\bibnamefont {Valvin}}, \bibinfo {author} {\bibfnamefont {J.~H.}\
  \bibnamefont {Edgar}}, \bibinfo {author} {\bibfnamefont {G.}~\bibnamefont
  {Cassabois}}, \ and\ \bibinfo {author} {\bibfnamefont {B.}~\bibnamefont
  {Gil}},\ }\href {https://doi.org/10.1038/nmat5048} {\bibfield  {journal}
  {\bibinfo  {journal} {Nature Materials}\ }\textbf {\bibinfo {volume} {17}}
  (\bibinfo {year} {2018})}\BibitemShut {NoStop}%
\bibitem [{\citenamefont {Sponza}\ \emph {et~al.}(2018)\citenamefont {Sponza},
  \citenamefont {Amara}, \citenamefont {Ducastelle}, \citenamefont {Loiseau},\
  and\ \citenamefont {Attaccalite}}]{Sponza2018}%
  \BibitemOpen
  \bibfield  {author} {\bibinfo {author} {\bibfnamefont {L.}~\bibnamefont
  {Sponza}}, \bibinfo {author} {\bibfnamefont {H.}~\bibnamefont {Amara}},
  \bibinfo {author} {\bibfnamefont {F.}~\bibnamefont {Ducastelle}}, \bibinfo
  {author} {\bibfnamefont {A.}~\bibnamefont {Loiseau}}, \ and\ \bibinfo
  {author} {\bibfnamefont {C.}~\bibnamefont {Attaccalite}},\ }\href {\doibase
  10.1103/PhysRevB.97.075121} {\bibfield  {journal} {\bibinfo  {journal} {Phys.
  Rev. B}\ }\textbf {\bibinfo {volume} {97}},\ \bibinfo {pages} {075121}
  (\bibinfo {year} {2018})}\BibitemShut {NoStop}%
\bibitem [{\citenamefont {Schu\'e}\ \emph {et~al.}(2019)\citenamefont
  {Schu\'e}, \citenamefont {Sponza}, \citenamefont {Plaud}, \citenamefont
  {Bensalah}, \citenamefont {Watanabe}, \citenamefont {Taniguchi},
  \citenamefont {Ducastelle}, \citenamefont {Loiseau},\ and\ \citenamefont
  {Barjon}}]{Schue2019}%
  \BibitemOpen
  \bibfield  {author} {\bibinfo {author} {\bibfnamefont {L.}~\bibnamefont
  {Schu\'e}}, \bibinfo {author} {\bibfnamefont {L.}~\bibnamefont {Sponza}},
  \bibinfo {author} {\bibfnamefont {A.}~\bibnamefont {Plaud}}, \bibinfo
  {author} {\bibfnamefont {H.}~\bibnamefont {Bensalah}}, \bibinfo {author}
  {\bibfnamefont {K.}~\bibnamefont {Watanabe}}, \bibinfo {author}
  {\bibfnamefont {T.}~\bibnamefont {Taniguchi}}, \bibinfo {author}
  {\bibfnamefont {F.}~\bibnamefont {Ducastelle}}, \bibinfo {author}
  {\bibfnamefont {A.}~\bibnamefont {Loiseau}}, \ and\ \bibinfo {author}
  {\bibfnamefont {J.}~\bibnamefont {Barjon}},\ }\href {\doibase
  10.1103/PhysRevLett.122.067401} {\bibfield  {journal} {\bibinfo  {journal}
  {Phys. Rev. Lett.}\ }\textbf {\bibinfo {volume} {122}},\ \bibinfo {pages}
  {067401} (\bibinfo {year} {2019})}\BibitemShut {NoStop}%
\bibitem [{\citenamefont {Cannuccia}\ \emph {et~al.}(2019)\citenamefont
  {Cannuccia}, \citenamefont {Monserrat},\ and\ \citenamefont
  {Attaccalite}}]{Cannuccia2019}%
  \BibitemOpen
  \bibfield  {author} {\bibinfo {author} {\bibfnamefont {E.}~\bibnamefont
  {Cannuccia}}, \bibinfo {author} {\bibfnamefont {B.}~\bibnamefont
  {Monserrat}}, \ and\ \bibinfo {author} {\bibfnamefont {C.}~\bibnamefont
  {Attaccalite}},\ }\href {\doibase 10.1103/PhysRevB.99.081109} {\bibfield
  {journal} {\bibinfo  {journal} {Phys. Rev. B}\ }\textbf {\bibinfo {volume}
  {99}},\ \bibinfo {pages} {081109} (\bibinfo {year} {2019})}\BibitemShut
  {NoStop}%
\bibitem [{\citenamefont {Paleari}\ \emph {et~al.}(2019)\citenamefont
  {Paleari}, \citenamefont {P.C.Miranda}, \citenamefont {Molina-S\'anchez},\
  and\ \citenamefont {Wirtz}}]{Paleari2019}%
  \BibitemOpen
  \bibfield  {author} {\bibinfo {author} {\bibfnamefont {F.}~\bibnamefont
  {Paleari}}, \bibinfo {author} {\bibfnamefont {H.}~\bibnamefont
  {P.C.Miranda}}, \bibinfo {author} {\bibfnamefont {A.}~\bibnamefont
  {Molina-S\'anchez}}, \ and\ \bibinfo {author} {\bibfnamefont
  {L.}~\bibnamefont {Wirtz}},\ }\href {\doibase 10.1103/PhysRevLett.122.187401}
  {\bibfield  {journal} {\bibinfo  {journal} {Phys. Rev. Lett.}\ }\textbf
  {\bibinfo {volume} {122}},\ \bibinfo {pages} {187401} (\bibinfo {year}
  {2019})}\BibitemShut {NoStop}%
\bibitem [{\citenamefont {Roux}\ \emph {et~al.}(2021)\citenamefont {Roux},
  \citenamefont {Arnold}, \citenamefont {Paleari}, \citenamefont {Sponza},
  \citenamefont {Janzen}, \citenamefont {Edgar}, \citenamefont {Toury},
  \citenamefont {Journet}, \citenamefont {Garnier}, \citenamefont {Steyer},
  \citenamefont {Taniguchi}, \citenamefont {Watanabe}, \citenamefont
  {Ducastelle}, \citenamefont {Loiseau},\ and\ \citenamefont
  {Barjon}}]{Roux2021}%
  \BibitemOpen
  \bibfield  {author} {\bibinfo {author} {\bibfnamefont {S.}~\bibnamefont
  {Roux}}, \bibinfo {author} {\bibfnamefont {C.}~\bibnamefont {Arnold}},
  \bibinfo {author} {\bibfnamefont {F.}~\bibnamefont {Paleari}}, \bibinfo
  {author} {\bibfnamefont {L.}~\bibnamefont {Sponza}}, \bibinfo {author}
  {\bibfnamefont {E.}~\bibnamefont {Janzen}}, \bibinfo {author} {\bibfnamefont
  {J.~H.}\ \bibnamefont {Edgar}}, \bibinfo {author} {\bibfnamefont
  {B.}~\bibnamefont {Toury}}, \bibinfo {author} {\bibfnamefont
  {C.}~\bibnamefont {Journet}}, \bibinfo {author} {\bibfnamefont
  {V.}~\bibnamefont {Garnier}}, \bibinfo {author} {\bibfnamefont
  {P.}~\bibnamefont {Steyer}}, \bibinfo {author} {\bibfnamefont
  {T.}~\bibnamefont {Taniguchi}}, \bibinfo {author} {\bibfnamefont
  {K.}~\bibnamefont {Watanabe}}, \bibinfo {author} {\bibfnamefont
  {F.}~\bibnamefont {Ducastelle}}, \bibinfo {author} {\bibfnamefont
  {A.}~\bibnamefont {Loiseau}}, \ and\ \bibinfo {author} {\bibfnamefont
  {J.}~\bibnamefont {Barjon}},\ }\href {\doibase 10.1103/PhysRevB.104.L161203}
  {\bibfield  {journal} {\bibinfo  {journal} {Phys. Rev. B}\ }\textbf {\bibinfo
  {volume} {104}},\ \bibinfo {pages} {L161203} (\bibinfo {year}
  {2021})}\BibitemShut {NoStop}%
\bibitem [{\citenamefont {Nagakubo}\ \emph {et~al.}(2013)\citenamefont
  {Nagakubo}, \citenamefont {Ogi}, \citenamefont {Sumiya}, \citenamefont
  {Kusakabe},\ and\ \citenamefont {Hirao}}]{Nagakubo2013}%
  \BibitemOpen
  \bibfield  {author} {\bibinfo {author} {\bibfnamefont {A.}~\bibnamefont
  {Nagakubo}}, \bibinfo {author} {\bibfnamefont {H.}~\bibnamefont {Ogi}},
  \bibinfo {author} {\bibfnamefont {H.}~\bibnamefont {Sumiya}}, \bibinfo
  {author} {\bibfnamefont {K.}~\bibnamefont {Kusakabe}}, \ and\ \bibinfo
  {author} {\bibfnamefont {M.}~\bibnamefont {Hirao}},\ }\href {\doibase
  10.1063/1.4811789} {\bibfield  {journal} {\bibinfo  {journal} {Applied
  Physics Letters}\ }\textbf {\bibinfo {volume} {102}},\ \bibinfo {pages}
  {241909} (\bibinfo {year} {2013})}\BibitemShut {NoStop}%
\bibitem [{\citenamefont {Deura}\ \emph {et~al.}(2017)\citenamefont {Deura},
  \citenamefont {Kutsukake}, \citenamefont {Ohno}, \citenamefont {Yonenaga},\
  and\ \citenamefont {Taniguchi}}]{Deura2017}%
  \BibitemOpen
  \bibfield  {author} {\bibinfo {author} {\bibfnamefont {M.}~\bibnamefont
  {Deura}}, \bibinfo {author} {\bibfnamefont {K.}~\bibnamefont {Kutsukake}},
  \bibinfo {author} {\bibfnamefont {Y.}~\bibnamefont {Ohno}}, \bibinfo {author}
  {\bibfnamefont {I.}~\bibnamefont {Yonenaga}}, \ and\ \bibinfo {author}
  {\bibfnamefont {T.}~\bibnamefont {Taniguchi}},\ }\href {\doibase
  10.7567/jjap.56.030301} {\bibfield  {journal} {\bibinfo  {journal} {Japanese
  Journal of Applied Physics}\ }\textbf {\bibinfo {volume} {56}},\ \bibinfo
  {pages} {030301} (\bibinfo {year} {2017})}\BibitemShut {NoStop}%
\bibitem [{\citenamefont {Dreyer}\ \emph {et~al.}(2014)\citenamefont {Dreyer},
  \citenamefont {Lyons}, \citenamefont {Janotti},\ and\ \citenamefont
  {de~Walle}}]{Dreyer2014}%
  \BibitemOpen
  \bibfield  {author} {\bibinfo {author} {\bibfnamefont {C.~E.}\ \bibnamefont
  {Dreyer}}, \bibinfo {author} {\bibfnamefont {J.~L.}\ \bibnamefont {Lyons}},
  \bibinfo {author} {\bibfnamefont {A.}~\bibnamefont {Janotti}}, \ and\
  \bibinfo {author} {\bibfnamefont {C.~G.~V.}\ \bibnamefont {de~Walle}},\
  }\href {\doibase 10.7567/APEX.7.031001} {\bibfield  {journal} {\bibinfo
  {journal} {Applied Physics Express}\ }\textbf {\bibinfo {volume} {7}},\
  \bibinfo {pages} {031001} (\bibinfo {year} {2014})}\BibitemShut {NoStop}%
\bibitem [{\citenamefont {Yixi}\ \emph {et~al.}(1994)\citenamefont {Yixi},
  \citenamefont {Xin}, \citenamefont {Kun}, \citenamefont {Chaoshu},
  \citenamefont {Zhengfu}, \citenamefont {Junyan}, \citenamefont {Jie},
  \citenamefont {Sheng},\ and\ \citenamefont {Yuanbin}}]{Yixi1994}%
  \BibitemOpen
  \bibfield  {author} {\bibinfo {author} {\bibfnamefont {S.}~\bibnamefont
  {Yixi}}, \bibinfo {author} {\bibfnamefont {J.}~\bibnamefont {Xin}}, \bibinfo
  {author} {\bibfnamefont {W.}~\bibnamefont {Kun}}, \bibinfo {author}
  {\bibfnamefont {S.}~\bibnamefont {Chaoshu}}, \bibinfo {author} {\bibfnamefont
  {H.}~\bibnamefont {Zhengfu}}, \bibinfo {author} {\bibfnamefont
  {S.}~\bibnamefont {Junyan}}, \bibinfo {author} {\bibfnamefont
  {D.}~\bibnamefont {Jie}}, \bibinfo {author} {\bibfnamefont {Z.}~\bibnamefont
  {Sheng}}, \ and\ \bibinfo {author} {\bibfnamefont {C.}~\bibnamefont
  {Yuanbin}},\ }\href {\doibase 10.1103/PhysRevB.50.18637} {\bibfield
  {journal} {\bibinfo  {journal} {Phys. Rev. B}\ }\textbf {\bibinfo {volume}
  {50}},\ \bibinfo {pages} {18637} (\bibinfo {year} {1994})}\BibitemShut
  {NoStop}%
\bibitem [{\citenamefont {Liu}\ \emph {et~al.}(2019)\citenamefont {Liu},
  \citenamefont {Wang}, \citenamefont {He}, \citenamefont {Zhang},
  \citenamefont {Liang}, \citenamefont {Moellendick}, \citenamefont {Zhao},\
  and\ \citenamefont {Li}}]{Liu2019}%
  \BibitemOpen
  \bibfield  {author} {\bibinfo {author} {\bibfnamefont {Y.}~\bibnamefont
  {Liu}}, \bibinfo {author} {\bibfnamefont {Q.}~\bibnamefont {Wang}}, \bibinfo
  {author} {\bibfnamefont {D.}~\bibnamefont {He}}, \bibinfo {author}
  {\bibfnamefont {J.}~\bibnamefont {Zhang}}, \bibinfo {author} {\bibfnamefont
  {A.}~\bibnamefont {Liang}}, \bibinfo {author} {\bibfnamefont {T.~E.}\
  \bibnamefont {Moellendick}}, \bibinfo {author} {\bibfnamefont
  {L.}~\bibnamefont {Zhao}}, \ and\ \bibinfo {author} {\bibfnamefont
  {X.}~\bibnamefont {Li}},\ }\href {https://doi.org/10.1038/s41598-019-46709-4}
  {\bibfield  {journal} {\bibinfo  {journal} {Scientific Reports}\ }\textbf
  {\bibinfo {volume} {9}} (\bibinfo {year} {2019})}\BibitemShut {NoStop}%
\bibitem [{\citenamefont {Segura}\ \emph {et~al.}(2019)\citenamefont {Segura},
  \citenamefont {Cuscó}, \citenamefont {Taniguchi}, \citenamefont {Watanabe},
  \citenamefont {Cassabois}, \citenamefont {Gil},\ and\ \citenamefont
  {Artús}}]{Segura2019-nonreversiblewbn}%
  \BibitemOpen
  \bibfield  {author} {\bibinfo {author} {\bibfnamefont {A.}~\bibnamefont
  {Segura}}, \bibinfo {author} {\bibfnamefont {R.}~\bibnamefont {Cuscó}},
  \bibinfo {author} {\bibfnamefont {T.}~\bibnamefont {Taniguchi}}, \bibinfo
  {author} {\bibfnamefont {K.}~\bibnamefont {Watanabe}}, \bibinfo {author}
  {\bibfnamefont {G.}~\bibnamefont {Cassabois}}, \bibinfo {author}
  {\bibfnamefont {B.}~\bibnamefont {Gil}}, \ and\ \bibinfo {author}
  {\bibfnamefont {L.}~\bibnamefont {Artús}},\ }\href
  {https://doi.org/10.1021/acs.jpcc.9b06163} {\bibfield  {journal} {\bibinfo
  {journal} {The Journal of Physical Chemistry C}\ }\textbf {\bibinfo {volume}
  {123}},\ \bibinfo {pages} {20167} (\bibinfo {year} {2019})}\BibitemShut
  {NoStop}%
\bibitem [{\citenamefont {Chen}\ \emph {et~al.}(2019)\citenamefont {Chen},
  \citenamefont {Yin}, \citenamefont {Kato}, \citenamefont {Taniguchi},
  \citenamefont {Watanabe}, \citenamefont {Ma}, \citenamefont {Ye},\ and\
  \citenamefont {Ikuhara}}]{Chen2019}%
  \BibitemOpen
  \bibfield  {author} {\bibinfo {author} {\bibfnamefont {C.}~\bibnamefont
  {Chen}}, \bibinfo {author} {\bibfnamefont {D.}~\bibnamefont {Yin}}, \bibinfo
  {author} {\bibfnamefont {T.}~\bibnamefont {Kato}}, \bibinfo {author}
  {\bibfnamefont {T.}~\bibnamefont {Taniguchi}}, \bibinfo {author}
  {\bibfnamefont {K.}~\bibnamefont {Watanabe}}, \bibinfo {author}
  {\bibfnamefont {X.}~\bibnamefont {Ma}}, \bibinfo {author} {\bibfnamefont
  {H.}~\bibnamefont {Ye}}, \ and\ \bibinfo {author} {\bibfnamefont
  {Y.}~\bibnamefont {Ikuhara}},\ }\href {\doibase 10.1073/pnas.1902820116}
  {\bibfield  {journal} {\bibinfo  {journal} {Proceedings of the National
  Academy of Sciences}\ }\textbf {\bibinfo {volume} {116}},\ \bibinfo {pages}
  {11181} (\bibinfo {year} {2019})}\BibitemShut {NoStop}%
\bibitem [{\citenamefont {Tararan}\ \emph {et~al.}(2018)\citenamefont
  {Tararan}, \citenamefont {di~Sabatino}, \citenamefont {Gatti}, \citenamefont
  {Taniguchi}, \citenamefont {Watanabe}, \citenamefont {Reining}, \citenamefont
  {Tizei}, \citenamefont {Kociak},\ and\ \citenamefont
  {Zobelli}}]{Tararan2018}%
  \BibitemOpen
  \bibfield  {author} {\bibinfo {author} {\bibfnamefont {A.}~\bibnamefont
  {Tararan}}, \bibinfo {author} {\bibfnamefont {S.}~\bibnamefont
  {di~Sabatino}}, \bibinfo {author} {\bibfnamefont {M.}~\bibnamefont {Gatti}},
  \bibinfo {author} {\bibfnamefont {T.}~\bibnamefont {Taniguchi}}, \bibinfo
  {author} {\bibfnamefont {K.}~\bibnamefont {Watanabe}}, \bibinfo {author}
  {\bibfnamefont {L.}~\bibnamefont {Reining}}, \bibinfo {author} {\bibfnamefont
  {L.~H.~G.}\ \bibnamefont {Tizei}}, \bibinfo {author} {\bibfnamefont
  {M.}~\bibnamefont {Kociak}}, \ and\ \bibinfo {author} {\bibfnamefont
  {A.}~\bibnamefont {Zobelli}},\ }\href {\doibase 10.1103/PhysRevB.98.094106}
  {\bibfield  {journal} {\bibinfo  {journal} {Phys. Rev. B}\ }\textbf {\bibinfo
  {volume} {98}},\ \bibinfo {pages} {094106} (\bibinfo {year}
  {2018})}\BibitemShut {NoStop}%
\bibitem [{\citenamefont {Mishima}\ and\ \citenamefont
  {Era}(2000)}]{Mishima2000}%
  \BibitemOpen
  \bibfield  {author} {\bibinfo {author} {\bibfnamefont {O.}~\bibnamefont
  {Mishima}}\ and\ \bibinfo {author} {\bibfnamefont {K.}~\bibnamefont {Era}},\
  }\href {\doibase https://doi.org/10.1201/9780203908181} {\emph {\bibinfo
  {title} {Electric refractory materials}}},\ edited by\ \bibinfo {editor}
  {\bibfnamefont {Y.}~\bibnamefont {Kumashiro}}\ (\bibinfo  {publisher} {Marcel
  Dekker, New York},\ \bibinfo {year} {2000})\ Chap.~\bibinfo {chapter}
  {21}\BibitemShut {NoStop}%
\bibitem [{\citenamefont {Horiuchi}\ \emph {et~al.}(1996)\citenamefont
  {Horiuchi}, \citenamefont {He}, \citenamefont {Onoda},\ and\ \citenamefont
  {Akaishi}}]{Horiuchi1996}%
  \BibitemOpen
  \bibfield  {author} {\bibinfo {author} {\bibfnamefont {S.}~\bibnamefont
  {Horiuchi}}, \bibinfo {author} {\bibfnamefont {L.}~\bibnamefont {He}},
  \bibinfo {author} {\bibfnamefont {M.}~\bibnamefont {Onoda}}, \ and\ \bibinfo
  {author} {\bibfnamefont {M.}~\bibnamefont {Akaishi}},\ }\href
  {https://doi.org/10.1063/1.116453} {\bibfield  {journal} {\bibinfo  {journal}
  {Applied Physics Letters}\ }\textbf {\bibinfo {volume} {68}},\ \bibinfo
  {pages} {182} (\bibinfo {year} {1996})}\BibitemShut {NoStop}%
\bibitem [{\citenamefont {Giannozzi}\ \emph {et~al.}(2009)\citenamefont
  {Giannozzi}, \citenamefont {Baroni}, \citenamefont {Bonini}, \citenamefont
  {Calandra}, \citenamefont {Car}, \citenamefont {Cavazzoni}, \citenamefont
  {Ceresoli}, \citenamefont {Chiarotti}, \citenamefont {Cococcioni},
  \citenamefont {Dabo}, \citenamefont {Corso}, \citenamefont {de~Gironcoli},
  \citenamefont {Fabris}, \citenamefont {Fratesi}, \citenamefont {Gebauer},
  \citenamefont {Gerstmann}, \citenamefont {Gougoussis}, \citenamefont
  {Kokalj}, \citenamefont {Lazzeri}, \citenamefont {Martin-Samos},
  \citenamefont {Marzari}, \citenamefont {Mauri}, \citenamefont {Mazzarello},
  \citenamefont {Paolini}, \citenamefont {Pasquarello}, \citenamefont
  {Paulatto}, \citenamefont {Sbraccia}, \citenamefont {Scandolo}, \citenamefont
  {Sclauzero}, \citenamefont {Seitsonen}, \citenamefont {Smogunov},
  \citenamefont {Umari},\ and\ \citenamefont {Wentzcovitch}}]{qe-code}%
  \BibitemOpen
  \bibfield  {author} {\bibinfo {author} {\bibfnamefont {P.}~\bibnamefont
  {Giannozzi}}, \bibinfo {author} {\bibfnamefont {S.}~\bibnamefont {Baroni}},
  \bibinfo {author} {\bibfnamefont {N.}~\bibnamefont {Bonini}}, \bibinfo
  {author} {\bibfnamefont {M.}~\bibnamefont {Calandra}}, \bibinfo {author}
  {\bibfnamefont {R.}~\bibnamefont {Car}}, \bibinfo {author} {\bibfnamefont
  {C.}~\bibnamefont {Cavazzoni}}, \bibinfo {author} {\bibfnamefont
  {D.}~\bibnamefont {Ceresoli}}, \bibinfo {author} {\bibfnamefont {G.~L.}\
  \bibnamefont {Chiarotti}}, \bibinfo {author} {\bibfnamefont {M.}~\bibnamefont
  {Cococcioni}}, \bibinfo {author} {\bibfnamefont {I.}~\bibnamefont {Dabo}},
  \bibinfo {author} {\bibfnamefont {A.~D.}\ \bibnamefont {Corso}}, \bibinfo
  {author} {\bibfnamefont {S.}~\bibnamefont {de~Gironcoli}}, \bibinfo {author}
  {\bibfnamefont {S.}~\bibnamefont {Fabris}}, \bibinfo {author} {\bibfnamefont
  {G.}~\bibnamefont {Fratesi}}, \bibinfo {author} {\bibfnamefont
  {R.}~\bibnamefont {Gebauer}}, \bibinfo {author} {\bibfnamefont
  {U.}~\bibnamefont {Gerstmann}}, \bibinfo {author} {\bibfnamefont
  {C.}~\bibnamefont {Gougoussis}}, \bibinfo {author} {\bibfnamefont
  {A.}~\bibnamefont {Kokalj}}, \bibinfo {author} {\bibfnamefont
  {M.}~\bibnamefont {Lazzeri}}, \bibinfo {author} {\bibfnamefont
  {L.}~\bibnamefont {Martin-Samos}}, \bibinfo {author} {\bibfnamefont
  {N.}~\bibnamefont {Marzari}}, \bibinfo {author} {\bibfnamefont
  {F.}~\bibnamefont {Mauri}}, \bibinfo {author} {\bibfnamefont
  {R.}~\bibnamefont {Mazzarello}}, \bibinfo {author} {\bibfnamefont
  {S.}~\bibnamefont {Paolini}}, \bibinfo {author} {\bibfnamefont
  {A.}~\bibnamefont {Pasquarello}}, \bibinfo {author} {\bibfnamefont
  {L.}~\bibnamefont {Paulatto}}, \bibinfo {author} {\bibfnamefont
  {C.}~\bibnamefont {Sbraccia}}, \bibinfo {author} {\bibfnamefont
  {S.}~\bibnamefont {Scandolo}}, \bibinfo {author} {\bibfnamefont
  {G.}~\bibnamefont {Sclauzero}}, \bibinfo {author} {\bibfnamefont {A.~P.}\
  \bibnamefont {Seitsonen}}, \bibinfo {author} {\bibfnamefont {A.}~\bibnamefont
  {Smogunov}}, \bibinfo {author} {\bibfnamefont {P.}~\bibnamefont {Umari}}, \
  and\ \bibinfo {author} {\bibfnamefont {R.~M.}\ \bibnamefont {Wentzcovitch}},\
  }\href {\doibase 10.1088/0953-8984/21/39/395502} {\bibfield  {journal}
  {\bibinfo  {journal} {Journal of Physics: Condensed Matter}\ }\textbf
  {\bibinfo {volume} {21}},\ \bibinfo {pages} {395502} (\bibinfo {year}
  {2009})}\BibitemShut {NoStop}%
\bibitem [{\citenamefont {Giannozzi}\ \emph {et~al.}(2017)\citenamefont
  {Giannozzi}, \citenamefont {Andreussi}, \citenamefont {Brumme}, \citenamefont
  {Bunau}, \citenamefont {Nardelli}, \citenamefont {Calandra}, \citenamefont
  {Car}, \citenamefont {Cavazzoni}, \citenamefont {Ceresoli}, \citenamefont
  {Cococcioni}, \citenamefont {Colonna}, \citenamefont {Carnimeo},
  \citenamefont {Corso}, \citenamefont {de~Gironcoli}, \citenamefont {Delugas},
  \citenamefont {DiStasio}, \citenamefont {Ferretti}, \citenamefont {Floris},
  \citenamefont {Fratesi}, \citenamefont {Fugallo}, \citenamefont {Gebauer},
  \citenamefont {Gerstmann}, \citenamefont {Giustino}, \citenamefont {Gorni},
  \citenamefont {Jia}, \citenamefont {Kawamura}, \citenamefont {Ko},
  \citenamefont {Kokalj}, \citenamefont {K\"u{\c{c}}kbenli}, \citenamefont
  {Lazzeri}, \citenamefont {Marsili}, \citenamefont {Marzari}, \citenamefont
  {Mauri}, \citenamefont {Nguyen}, \citenamefont {Nguyen}, \citenamefont {de-la
  Roza}, \citenamefont {Paulatto}, \citenamefont {Ponc{\'{e}}}, \citenamefont
  {Rocca}, \citenamefont {Sabatini}, \citenamefont {Santra}, \citenamefont
  {Schlipf}, \citenamefont {Seitsonen}, \citenamefont {Smogunov}, \citenamefont
  {Timrov}, \citenamefont {Thonhauser}, \citenamefont {Umari}, \citenamefont
  {Vast}, \citenamefont {Wu},\ and\ \citenamefont {Baroni}}]{Giannozzi_2017}%
  \BibitemOpen
  \bibfield  {author} {\bibinfo {author} {\bibfnamefont {P.}~\bibnamefont
  {Giannozzi}}, \bibinfo {author} {\bibfnamefont {O.}~\bibnamefont
  {Andreussi}}, \bibinfo {author} {\bibfnamefont {T.}~\bibnamefont {Brumme}},
  \bibinfo {author} {\bibfnamefont {O.}~\bibnamefont {Bunau}}, \bibinfo
  {author} {\bibfnamefont {M.~B.}\ \bibnamefont {Nardelli}}, \bibinfo {author}
  {\bibfnamefont {M.}~\bibnamefont {Calandra}}, \bibinfo {author}
  {\bibfnamefont {R.}~\bibnamefont {Car}}, \bibinfo {author} {\bibfnamefont
  {C.}~\bibnamefont {Cavazzoni}}, \bibinfo {author} {\bibfnamefont
  {D.}~\bibnamefont {Ceresoli}}, \bibinfo {author} {\bibfnamefont
  {M.}~\bibnamefont {Cococcioni}}, \bibinfo {author} {\bibfnamefont
  {N.}~\bibnamefont {Colonna}}, \bibinfo {author} {\bibfnamefont
  {I.}~\bibnamefont {Carnimeo}}, \bibinfo {author} {\bibfnamefont {A.~D.}\
  \bibnamefont {Corso}}, \bibinfo {author} {\bibfnamefont {S.}~\bibnamefont
  {de~Gironcoli}}, \bibinfo {author} {\bibfnamefont {P.}~\bibnamefont
  {Delugas}}, \bibinfo {author} {\bibfnamefont {R.~A.}\ \bibnamefont
  {DiStasio}}, \bibinfo {author} {\bibfnamefont {A.}~\bibnamefont {Ferretti}},
  \bibinfo {author} {\bibfnamefont {A.}~\bibnamefont {Floris}}, \bibinfo
  {author} {\bibfnamefont {G.}~\bibnamefont {Fratesi}}, \bibinfo {author}
  {\bibfnamefont {G.}~\bibnamefont {Fugallo}}, \bibinfo {author} {\bibfnamefont
  {R.}~\bibnamefont {Gebauer}}, \bibinfo {author} {\bibfnamefont
  {U.}~\bibnamefont {Gerstmann}}, \bibinfo {author} {\bibfnamefont
  {F.}~\bibnamefont {Giustino}}, \bibinfo {author} {\bibfnamefont
  {T.}~\bibnamefont {Gorni}}, \bibinfo {author} {\bibfnamefont
  {J.}~\bibnamefont {Jia}}, \bibinfo {author} {\bibfnamefont {M.}~\bibnamefont
  {Kawamura}}, \bibinfo {author} {\bibfnamefont {H.-Y.}\ \bibnamefont {Ko}},
  \bibinfo {author} {\bibfnamefont {A.}~\bibnamefont {Kokalj}}, \bibinfo
  {author} {\bibfnamefont {E.}~\bibnamefont {K\"u{\c{c}}kbenli}}, \bibinfo
  {author} {\bibfnamefont {M.}~\bibnamefont {Lazzeri}}, \bibinfo {author}
  {\bibfnamefont {M.}~\bibnamefont {Marsili}}, \bibinfo {author} {\bibfnamefont
  {N.}~\bibnamefont {Marzari}}, \bibinfo {author} {\bibfnamefont
  {F.}~\bibnamefont {Mauri}}, \bibinfo {author} {\bibfnamefont {N.~L.}\
  \bibnamefont {Nguyen}}, \bibinfo {author} {\bibfnamefont {H.-V.}\
  \bibnamefont {Nguyen}}, \bibinfo {author} {\bibfnamefont {A.~O.}\
  \bibnamefont {de-la Roza}}, \bibinfo {author} {\bibfnamefont
  {L.}~\bibnamefont {Paulatto}}, \bibinfo {author} {\bibfnamefont
  {S.}~\bibnamefont {Ponc{\'{e}}}}, \bibinfo {author} {\bibfnamefont
  {D.}~\bibnamefont {Rocca}}, \bibinfo {author} {\bibfnamefont
  {R.}~\bibnamefont {Sabatini}}, \bibinfo {author} {\bibfnamefont
  {B.}~\bibnamefont {Santra}}, \bibinfo {author} {\bibfnamefont
  {M.}~\bibnamefont {Schlipf}}, \bibinfo {author} {\bibfnamefont {A.~P.}\
  \bibnamefont {Seitsonen}}, \bibinfo {author} {\bibfnamefont {A.}~\bibnamefont
  {Smogunov}}, \bibinfo {author} {\bibfnamefont {I.}~\bibnamefont {Timrov}},
  \bibinfo {author} {\bibfnamefont {T.}~\bibnamefont {Thonhauser}}, \bibinfo
  {author} {\bibfnamefont {P.}~\bibnamefont {Umari}}, \bibinfo {author}
  {\bibfnamefont {N.}~\bibnamefont {Vast}}, \bibinfo {author} {\bibfnamefont
  {X.}~\bibnamefont {Wu}}, \ and\ \bibinfo {author} {\bibfnamefont
  {S.}~\bibnamefont {Baroni}},\ }\href {\doibase 10.1088/1361-648x/aa8f79}
  {\bibfield  {journal} {\bibinfo  {journal} {Journal of Physics: Condensed
  Matter}\ }\textbf {\bibinfo {volume} {29}},\ \bibinfo {pages} {465901}
  (\bibinfo {year} {2017})}\BibitemShut {NoStop}%
\bibitem [{\citenamefont {Perdew}\ \emph {et~al.}(2008)\citenamefont {Perdew},
  \citenamefont {Ruzsinszky}, \citenamefont {Csonka}, \citenamefont {Vydrov},
  \citenamefont {Scuseria}, \citenamefont {Constantin}, \citenamefont {Zhou},\
  and\ \citenamefont {Burke}}]{perdew2008restoring}%
  \BibitemOpen
  \bibfield  {author} {\bibinfo {author} {\bibfnamefont {J.~P.}\ \bibnamefont
  {Perdew}}, \bibinfo {author} {\bibfnamefont {A.}~\bibnamefont {Ruzsinszky}},
  \bibinfo {author} {\bibfnamefont {G.~I.}\ \bibnamefont {Csonka}}, \bibinfo
  {author} {\bibfnamefont {O.~A.}\ \bibnamefont {Vydrov}}, \bibinfo {author}
  {\bibfnamefont {G.~E.}\ \bibnamefont {Scuseria}}, \bibinfo {author}
  {\bibfnamefont {L.~A.}\ \bibnamefont {Constantin}}, \bibinfo {author}
  {\bibfnamefont {X.}~\bibnamefont {Zhou}}, \ and\ \bibinfo {author}
  {\bibfnamefont {K.}~\bibnamefont {Burke}},\ }\href@noop {} {\bibfield
  {journal} {\bibinfo  {journal} {Physical review letters}\ }\textbf {\bibinfo
  {volume} {100}},\ \bibinfo {pages} {136406} (\bibinfo {year}
  {2008})}\BibitemShut {NoStop}%
\bibitem [{\citenamefont {van Setten}\ \emph {et~al.}(2018)\citenamefont {van
  Setten}, \citenamefont {Giantomassi}, \citenamefont {Bousquet}, \citenamefont
  {Verstraete}, \citenamefont {Hamann}, \citenamefont {Gonze},\ and\
  \citenamefont {Rignanese}}]{van2018pseudodojo}%
  \BibitemOpen
  \bibfield  {author} {\bibinfo {author} {\bibfnamefont {M.~J.}\ \bibnamefont
  {van Setten}}, \bibinfo {author} {\bibfnamefont {M.}~\bibnamefont
  {Giantomassi}}, \bibinfo {author} {\bibfnamefont {E.}~\bibnamefont
  {Bousquet}}, \bibinfo {author} {\bibfnamefont {M.~J.}\ \bibnamefont
  {Verstraete}}, \bibinfo {author} {\bibfnamefont {D.~R.}\ \bibnamefont
  {Hamann}}, \bibinfo {author} {\bibfnamefont {X.}~\bibnamefont {Gonze}}, \
  and\ \bibinfo {author} {\bibfnamefont {G.-M.}\ \bibnamefont {Rignanese}},\
  }\href@noop {} {\bibfield  {journal} {\bibinfo  {journal} {Computer Physics
  Communications}\ }\textbf {\bibinfo {volume} {226}},\ \bibinfo {pages} {39}
  (\bibinfo {year} {2018})}\BibitemShut {NoStop}%
\bibitem [{\citenamefont {Onida}\ \emph {et~al.}(2002)\citenamefont {Onida},
  \citenamefont {Reining},\ and\ \citenamefont {Rubio}}]{onida_reining_rubio}%
  \BibitemOpen
  \bibfield  {author} {\bibinfo {author} {\bibfnamefont {G.}~\bibnamefont
  {Onida}}, \bibinfo {author} {\bibfnamefont {L.}~\bibnamefont {Reining}}, \
  and\ \bibinfo {author} {\bibfnamefont {A.}~\bibnamefont {Rubio}},\ }\href
  {\doibase 10.1103/RevModPhys.74.601} {\bibfield  {journal} {\bibinfo
  {journal} {Rev. Mod. Phys.}\ }\textbf {\bibinfo {volume} {74}},\ \bibinfo
  {pages} {601} (\bibinfo {year} {2002})}\BibitemShut {NoStop}%
\bibitem [{\citenamefont {Sangalli}\ \emph {et~al.}(2019)\citenamefont
  {Sangalli}, \citenamefont {Ferretti}, \citenamefont {Miranda}, \citenamefont
  {Attaccalite}, \citenamefont {Marri}, \citenamefont {Cannuccia},
  \citenamefont {Melo}, \citenamefont {Marsili}, \citenamefont {Paleari},
  \citenamefont {Marrazzo} \emph {et~al.}}]{yambo-code}%
  \BibitemOpen
  \bibfield  {author} {\bibinfo {author} {\bibfnamefont {D.}~\bibnamefont
  {Sangalli}}, \bibinfo {author} {\bibfnamefont {A.}~\bibnamefont {Ferretti}},
  \bibinfo {author} {\bibfnamefont {H.}~\bibnamefont {Miranda}}, \bibinfo
  {author} {\bibfnamefont {C.}~\bibnamefont {Attaccalite}}, \bibinfo {author}
  {\bibfnamefont {I.}~\bibnamefont {Marri}}, \bibinfo {author} {\bibfnamefont
  {E.}~\bibnamefont {Cannuccia}}, \bibinfo {author} {\bibfnamefont
  {P.}~\bibnamefont {Melo}}, \bibinfo {author} {\bibfnamefont {M.}~\bibnamefont
  {Marsili}}, \bibinfo {author} {\bibfnamefont {F.}~\bibnamefont {Paleari}},
  \bibinfo {author} {\bibfnamefont {A.}~\bibnamefont {Marrazzo}},  \emph
  {et~al.},\ }\href {\doibase 10.1088/1361-648X/ab15d0} {\bibfield  {journal}
  {\bibinfo  {journal} {Journal of Physics: Condensed Matter}\ }\textbf
  {\bibinfo {volume} {31}},\ \bibinfo {pages} {325902} (\bibinfo {year}
  {2019})}\BibitemShut {NoStop}%
\bibitem [{\citenamefont {Strinati}(1988)}]{strinati}%
  \BibitemOpen
  \bibfield  {author} {\bibinfo {author} {\bibfnamefont {G.}~\bibnamefont
  {Strinati}},\ }\href {\doibase 10.1007/BF02725962} {\bibfield  {journal}
  {\bibinfo  {journal} {Riv. Nuovo Cimento}\ }\textbf {\bibinfo {volume}
  {11}},\ \bibinfo {pages} {1} (\bibinfo {year} {1988})}\BibitemShut {NoStop}%
\bibitem [{\citenamefont {Chen}\ \emph {et~al.}(2020)\citenamefont {Chen},
  \citenamefont {Sangalli},\ and\ \citenamefont {Bernardi}}]{Cheng2020}%
  \BibitemOpen
  \bibfield  {author} {\bibinfo {author} {\bibfnamefont {H.-Y.}\ \bibnamefont
  {Chen}}, \bibinfo {author} {\bibfnamefont {D.}~\bibnamefont {Sangalli}}, \
  and\ \bibinfo {author} {\bibfnamefont {M.}~\bibnamefont {Bernardi}},\ }\href
  {\doibase 10.1103/PhysRevLett.125.107401} {\bibfield  {journal} {\bibinfo
  {journal} {Phys. Rev. Lett.}\ }\textbf {\bibinfo {volume} {125}},\ \bibinfo
  {pages} {107401} (\bibinfo {year} {2020})}\BibitemShut {NoStop}%
\bibitem [{\citenamefont {van Roosbroeck}\ and\ \citenamefont
  {Shockley}(1954)}]{Roosbroeck1954}%
  \BibitemOpen
  \bibfield  {author} {\bibinfo {author} {\bibfnamefont {W.}~\bibnamefont {van
  Roosbroeck}}\ and\ \bibinfo {author} {\bibfnamefont {W.}~\bibnamefont
  {Shockley}},\ }\href {\doibase 10.1103/PhysRev.94.1558} {\bibfield  {journal}
  {\bibinfo  {journal} {Phys. Rev.}\ }\textbf {\bibinfo {volume} {94}},\
  \bibinfo {pages} {1558} (\bibinfo {year} {1954})}\BibitemShut {NoStop}%
\bibitem [{\citenamefont {Bebb}\ and\ \citenamefont
  {Williams}(1972)}]{Bebb1972}%
  \BibitemOpen
  \bibfield  {author} {\bibinfo {author} {\bibfnamefont {H.}~\bibnamefont
  {Bebb}}\ and\ \bibinfo {author} {\bibfnamefont {E.}~\bibnamefont
  {Williams}},\ }\href {\doibase https://doi.org/10.1016/S0080-8784(08)62345-5}
  {\emph {\bibinfo {title} {Chapter 4 Photoluminescence I: Theory}}},\ edited
  by\ \bibinfo {editor} {\bibfnamefont {R.}~\bibnamefont {Willardson}}\ and\
  \bibinfo {editor} {\bibfnamefont {A.~C.}\ \bibnamefont {Beer}},\ \bibinfo
  {series} {Semiconductors and Semimetals}, Vol.~\bibinfo {volume} {8}\
  (\bibinfo  {publisher} {Elsevier},\ \bibinfo {year} {1972})\ pp.\ \bibinfo
  {pages} {181--320}\BibitemShut {NoStop}%
\bibitem [{\citenamefont {Izyumskaya}\ \emph {et~al.}(2017)\citenamefont
  {Izyumskaya}, \citenamefont {Demchenko}, \citenamefont {Das}, \citenamefont
  {\"{O}zg\"{u}r}, \citenamefont {Avrutin},\ and\ \citenamefont
  {Morkoc}}]{Izyumskaya2017}%
  \BibitemOpen
  \bibfield  {author} {\bibinfo {author} {\bibfnamefont {N.}~\bibnamefont
  {Izyumskaya}}, \bibinfo {author} {\bibfnamefont {D.~O.}\ \bibnamefont
  {Demchenko}}, \bibinfo {author} {\bibfnamefont {S.}~\bibnamefont {Das}},
  \bibinfo {author} {\bibfnamefont {U.}~\bibnamefont {\"{O}zg\"{u}r}}, \bibinfo
  {author} {\bibfnamefont {V.}~\bibnamefont {Avrutin}}, \ and\ \bibinfo
  {author} {\bibfnamefont {H.}~\bibnamefont {Morkoc}},\ }\href {\doibase
  https://doi.org/10.1002/aelm.201600485} {\bibfield  {journal} {\bibinfo
  {journal} {Advanced Electronic Materials}\ }\textbf {\bibinfo {volume} {3}},\
  \bibinfo {pages} {1600485} (\bibinfo {year} {2017})}\BibitemShut {NoStop}%
\bibitem [{SM()}]{SM}%
  \BibitemOpen
  \href@noop {} {}\bibinfo {howpublished} {See Supplemental Material at [URL
  will be inserted by publisher] for further information on lattice parameters,
  band structure, Born effective charges and phonons, which includes
  Refs.\cite{Karch1997,Solozhenko1998}.}\BibitemShut {Stop}%
\bibitem [{\citenamefont {Aryasetiawan}\ and\ \citenamefont
  {Gunnarsson}(1998)}]{aryasetiawan1998gw}%
  \BibitemOpen
  \bibfield  {author} {\bibinfo {author} {\bibfnamefont {F.}~\bibnamefont
  {Aryasetiawan}}\ and\ \bibinfo {author} {\bibfnamefont {O.}~\bibnamefont
  {Gunnarsson}},\ }\href {\doibase 10.1088/0034-4885/61/3/002} {\bibfield
  {journal} {\bibinfo  {journal} {Reports on Progress in Physics}\ }\textbf
  {\bibinfo {volume} {61}},\ \bibinfo {pages} {237} (\bibinfo {year}
  {1998})}\BibitemShut {NoStop}%
\bibitem [{\citenamefont {Godby}\ and\ \citenamefont
  {Needs}(1989)}]{PhysRevLett.62.1169}%
  \BibitemOpen
  \bibfield  {author} {\bibinfo {author} {\bibfnamefont {R.~W.}\ \bibnamefont
  {Godby}}\ and\ \bibinfo {author} {\bibfnamefont {R.~J.}\ \bibnamefont
  {Needs}},\ }\href {\doibase 10.1103/PhysRevLett.62.1169} {\bibfield
  {journal} {\bibinfo  {journal} {Phys. Rev. Lett.}\ }\textbf {\bibinfo
  {volume} {62}},\ \bibinfo {pages} {1169} (\bibinfo {year}
  {1989})}\BibitemShut {NoStop}%
\bibitem [{\citenamefont {Lloyd-Williams}\ and\ \citenamefont
  {Monserrat}(2015)}]{lloyd2015lattice}%
  \BibitemOpen
  \bibfield  {author} {\bibinfo {author} {\bibfnamefont {J.~H.}\ \bibnamefont
  {Lloyd-Williams}}\ and\ \bibinfo {author} {\bibfnamefont {B.}~\bibnamefont
  {Monserrat}},\ }\href@noop {} {\bibfield  {journal} {\bibinfo  {journal}
  {Physical Review B}\ }\textbf {\bibinfo {volume} {92}},\ \bibinfo {pages}
  {184301} (\bibinfo {year} {2015})}\BibitemShut {NoStop}%
\bibitem [{\citenamefont {Lechifflart}\ \emph {et~al.}(2022)\citenamefont
  {Lechifflart}, \citenamefont {Paleari},\ and\ \citenamefont
  {Attaccalite}}]{lechifflart2022excitons}%
  \BibitemOpen
  \bibfield  {author} {\bibinfo {author} {\bibfnamefont {P.}~\bibnamefont
  {Lechifflart}}, \bibinfo {author} {\bibfnamefont {F.}~\bibnamefont
  {Paleari}}, \ and\ \bibinfo {author} {\bibfnamefont {C.}~\bibnamefont
  {Attaccalite}},\ }\href {\doibase 10.21468/SciPostPhys.12.5.145} {\bibfield
  {journal} {\bibinfo  {journal} {SciPost Physics}\ }\textbf {\bibinfo {volume}
  {12}},\ \bibinfo {pages} {145} (\bibinfo {year} {2022})}\BibitemShut
  {NoStop}%
\bibitem [{\citenamefont {Yambo-Py}()}]{yambopy}%
  \BibitemOpen
  \bibfield  {author} {\bibinfo {author} {\bibnamefont {Yambo-Py}},\ }\href
  {https://github.com/yambo-code/yambopy} {\enquote {\bibinfo {title} {These
  operations have been performed with the aid of yambopy, a python-based
  interface to yambo and quantum espresso},}\ }\BibitemShut {NoStop}%
\bibitem [{\citenamefont {Yambo-wiki}()}]{yambowiki}%
  \BibitemOpen
  \bibfield  {author} {\bibinfo {author} {\bibnamefont {Yambo-wiki}},\ }\href
  {https://www.yambo-code.eu/wiki/index.php/Phonon-assisted_luminescence_by_finite_atomic_displacements}
  {\enquote {\bibinfo {title} {Phonon-assisted luminescence by finite atomic
  displacements},}\ }\BibitemShut {NoStop}%
\bibitem [{\citenamefont {Gorczyca}\ \emph {et~al.}(1995)\citenamefont
  {Gorczyca}, \citenamefont {Christensen}, \citenamefont {PeltzeryBlanc\'a},\
  and\ \citenamefont {Rodriguez}}]{Gorczyca1995}%
  \BibitemOpen
  \bibfield  {author} {\bibinfo {author} {\bibfnamefont {I.}~\bibnamefont
  {Gorczyca}}, \bibinfo {author} {\bibfnamefont {N.~E.}\ \bibnamefont
  {Christensen}}, \bibinfo {author} {\bibfnamefont {E.~L.}\ \bibnamefont
  {PeltzeryBlanc\'a}}, \ and\ \bibinfo {author} {\bibfnamefont {C.~O.}\
  \bibnamefont {Rodriguez}},\ }\href {\doibase 10.1103/PhysRevB.51.11936}
  {\bibfield  {journal} {\bibinfo  {journal} {Phys. Rev. B}\ }\textbf {\bibinfo
  {volume} {51}},\ \bibinfo {pages} {11936} (\bibinfo {year}
  {1995})}\BibitemShut {NoStop}%
\bibitem [{\citenamefont {Bosak}\ and\ \citenamefont
  {Krisch}(2006)}]{BOSAK20061661}%
  \BibitemOpen
  \bibfield  {author} {\bibinfo {author} {\bibfnamefont {A.}~\bibnamefont
  {Bosak}}\ and\ \bibinfo {author} {\bibfnamefont {M.}~\bibnamefont {Krisch}},\
  }\href {\doibase https://doi.org/10.1016/j.radphyschem.2005.07.024}
  {\bibfield  {journal} {\bibinfo  {journal} {Radiation Physics and Chemistry}\
  }\textbf {\bibinfo {volume} {75}},\ \bibinfo {pages} {1661} (\bibinfo {year}
  {2006})}\BibitemShut {NoStop}%
\bibitem [{\citenamefont {Karch}\ and\ \citenamefont
  {Bechstedt}(1997)}]{Karch1997}%
  \BibitemOpen
  \bibfield  {author} {\bibinfo {author} {\bibfnamefont {K.}~\bibnamefont
  {Karch}}\ and\ \bibinfo {author} {\bibfnamefont {F.}~\bibnamefont
  {Bechstedt}},\ }\href {\doibase 10.1103/PhysRevB.56.7404} {\bibfield
  {journal} {\bibinfo  {journal} {Phys. Rev. B}\ }\textbf {\bibinfo {volume}
  {56}},\ \bibinfo {pages} {7404} (\bibinfo {year} {1997})}\BibitemShut
  {NoStop}%
\bibitem [{\citenamefont {Perlin}\ \emph {et~al.}(1999)\citenamefont {Perlin},
  \citenamefont {Suski}, \citenamefont {Ager}, \citenamefont {Conti},
  \citenamefont {Polian}, \citenamefont {Christensen}, \citenamefont
  {Gorczyca}, \citenamefont {Grzegory}, \citenamefont {Weber},\ and\
  \citenamefont {Haller}}]{Perlin1999}%
  \BibitemOpen
  \bibfield  {author} {\bibinfo {author} {\bibfnamefont {P.}~\bibnamefont
  {Perlin}}, \bibinfo {author} {\bibfnamefont {T.}~\bibnamefont {Suski}},
  \bibinfo {author} {\bibfnamefont {J.~W.}\ \bibnamefont {Ager}}, \bibinfo
  {author} {\bibfnamefont {G.}~\bibnamefont {Conti}}, \bibinfo {author}
  {\bibfnamefont {A.}~\bibnamefont {Polian}}, \bibinfo {author} {\bibfnamefont
  {N.~E.}\ \bibnamefont {Christensen}}, \bibinfo {author} {\bibfnamefont
  {I.}~\bibnamefont {Gorczyca}}, \bibinfo {author} {\bibfnamefont
  {I.}~\bibnamefont {Grzegory}}, \bibinfo {author} {\bibfnamefont {E.~R.}\
  \bibnamefont {Weber}}, \ and\ \bibinfo {author} {\bibfnamefont {E.~E.}\
  \bibnamefont {Haller}},\ }\href {\doibase 10.1103/PhysRevB.60.1480}
  {\bibfield  {journal} {\bibinfo  {journal} {Phys. Rev. B}\ }\textbf {\bibinfo
  {volume} {60}},\ \bibinfo {pages} {1480} (\bibinfo {year}
  {1999})}\BibitemShut {NoStop}%
\bibitem [{\citenamefont {Go\~ni}\ \emph {et~al.}(2001)\citenamefont {Go\~ni},
  \citenamefont {Siegle}, \citenamefont {Syassen}, \citenamefont {Thomsen},\
  and\ \citenamefont {Wagner}}]{Gogni2001}%
  \BibitemOpen
  \bibfield  {author} {\bibinfo {author} {\bibfnamefont {A.~R.}\ \bibnamefont
  {Go\~ni}}, \bibinfo {author} {\bibfnamefont {H.}~\bibnamefont {Siegle}},
  \bibinfo {author} {\bibfnamefont {K.}~\bibnamefont {Syassen}}, \bibinfo
  {author} {\bibfnamefont {C.}~\bibnamefont {Thomsen}}, \ and\ \bibinfo
  {author} {\bibfnamefont {J.-M.}\ \bibnamefont {Wagner}},\ }\href {\doibase
  10.1103/PhysRevB.64.035205} {\bibfield  {journal} {\bibinfo  {journal} {Phys.
  Rev. B}\ }\textbf {\bibinfo {volume} {64}},\ \bibinfo {pages} {035205}
  (\bibinfo {year} {2001})}\BibitemShut {NoStop}%
\bibitem [{\citenamefont {Reparaz}\ \emph {et~al.}(2018)\citenamefont
  {Reparaz}, \citenamefont {da~Silva}, \citenamefont {Romero}, \citenamefont
  {Serrano}, \citenamefont {Wagner}, \citenamefont {Callsen}, \citenamefont
  {Choi}, \citenamefont {Speck},\ and\ \citenamefont {Go\~ni}}]{Reparaz2018}%
  \BibitemOpen
  \bibfield  {author} {\bibinfo {author} {\bibfnamefont {J.~S.}\ \bibnamefont
  {Reparaz}}, \bibinfo {author} {\bibfnamefont {K.~P.}\ \bibnamefont
  {da~Silva}}, \bibinfo {author} {\bibfnamefont {A.~H.}\ \bibnamefont
  {Romero}}, \bibinfo {author} {\bibfnamefont {J.}~\bibnamefont {Serrano}},
  \bibinfo {author} {\bibfnamefont {M.~R.}\ \bibnamefont {Wagner}}, \bibinfo
  {author} {\bibfnamefont {G.}~\bibnamefont {Callsen}}, \bibinfo {author}
  {\bibfnamefont {S.~J.}\ \bibnamefont {Choi}}, \bibinfo {author}
  {\bibfnamefont {J.~S.}\ \bibnamefont {Speck}}, \ and\ \bibinfo {author}
  {\bibfnamefont {A.~R.}\ \bibnamefont {Go\~ni}},\ }\href {\doibase
  10.1103/PhysRevB.98.165204} {\bibfield  {journal} {\bibinfo  {journal} {Phys.
  Rev. B}\ }\textbf {\bibinfo {volume} {98}},\ \bibinfo {pages} {165204}
  (\bibinfo {year} {2018})}\BibitemShut {NoStop}%
\bibitem [{\citenamefont {Loudon}(1964)}]{loudon1964}%
  \BibitemOpen
  \bibfield  {author} {\bibinfo {author} {\bibfnamefont {R.}~\bibnamefont
  {Loudon}},\ }\href {\doibase 10.1080/00018736400101051} {\bibfield  {journal}
  {\bibinfo  {journal} {Advances in Physics}\ }\textbf {\bibinfo {volume}
  {13}},\ \bibinfo {pages} {423} (\bibinfo {year} {1964})}\BibitemShut
  {NoStop}%
\bibitem [{\citenamefont {Feldman}\ \emph {et~al.}(1968)\citenamefont
  {Feldman}, \citenamefont {Parker}, \citenamefont {Choyke},\ and\
  \citenamefont {Patrick}}]{Feldman1968}%
  \BibitemOpen
  \bibfield  {author} {\bibinfo {author} {\bibfnamefont {D.~W.}\ \bibnamefont
  {Feldman}}, \bibinfo {author} {\bibfnamefont {J.~H.}\ \bibnamefont {Parker}},
  \bibinfo {author} {\bibfnamefont {W.~J.}\ \bibnamefont {Choyke}}, \ and\
  \bibinfo {author} {\bibfnamefont {L.}~\bibnamefont {Patrick}},\ }\href
  {\doibase 10.1103/PhysRev.173.787} {\bibfield  {journal} {\bibinfo  {journal}
  {Phys. Rev.}\ }\textbf {\bibinfo {volume} {173}},\ \bibinfo {pages} {787}
  (\bibinfo {year} {1968})}\BibitemShut {NoStop}%
\bibitem [{\citenamefont {Christensen}\ and\ \citenamefont
  {Gorczyca}(1994)}]{Christensen1994}%
  \BibitemOpen
  \bibfield  {author} {\bibinfo {author} {\bibfnamefont {N.~E.}\ \bibnamefont
  {Christensen}}\ and\ \bibinfo {author} {\bibfnamefont {I.}~\bibnamefont
  {Gorczyca}},\ }\href {\doibase 10.1103/PhysRevB.50.4397} {\bibfield
  {journal} {\bibinfo  {journal} {Phys. Rev. B}\ }\textbf {\bibinfo {volume}
  {50}},\ \bibinfo {pages} {4397} (\bibinfo {year} {1994})}\BibitemShut
  {NoStop}%
\bibitem [{\citenamefont {Kawai}\ \emph {et~al.}(2014)\citenamefont {Kawai},
  \citenamefont {Yamashita}, \citenamefont {Cannuccia},\ and\ \citenamefont
  {Marini}}]{kawai2014electron}%
  \BibitemOpen
  \bibfield  {author} {\bibinfo {author} {\bibfnamefont {H.}~\bibnamefont
  {Kawai}}, \bibinfo {author} {\bibfnamefont {K.}~\bibnamefont {Yamashita}},
  \bibinfo {author} {\bibfnamefont {E.}~\bibnamefont {Cannuccia}}, \ and\
  \bibinfo {author} {\bibfnamefont {A.}~\bibnamefont {Marini}},\ }\href
  {\doibase 10.1103/PhysRevB.89.085202} {\bibfield  {journal} {\bibinfo
  {journal} {Physical Review B}\ }\textbf {\bibinfo {volume} {89}},\ \bibinfo
  {pages} {085202} (\bibinfo {year} {2014})}\BibitemShut {NoStop}%
\bibitem [{\citenamefont {Elias}\ \emph {et~al.}(2021)\citenamefont {Elias},
  \citenamefont {Fugallo}, \citenamefont {Valvin}, \citenamefont {L’Henoret},
  \citenamefont {Li}, \citenamefont {Edgar}, \citenamefont {Sottile},
  \citenamefont {Lazzeri}, \citenamefont {Ouerghi}, \citenamefont {Gil} \emph
  {et~al.}}]{elias2021flat}%
  \BibitemOpen
  \bibfield  {author} {\bibinfo {author} {\bibfnamefont {C.}~\bibnamefont
  {Elias}}, \bibinfo {author} {\bibfnamefont {G.}~\bibnamefont {Fugallo}},
  \bibinfo {author} {\bibfnamefont {P.}~\bibnamefont {Valvin}}, \bibinfo
  {author} {\bibfnamefont {C.}~\bibnamefont {L’Henoret}}, \bibinfo {author}
  {\bibfnamefont {J.}~\bibnamefont {Li}}, \bibinfo {author} {\bibfnamefont
  {J.}~\bibnamefont {Edgar}}, \bibinfo {author} {\bibfnamefont
  {F.}~\bibnamefont {Sottile}}, \bibinfo {author} {\bibfnamefont
  {M.}~\bibnamefont {Lazzeri}}, \bibinfo {author} {\bibfnamefont
  {A.}~\bibnamefont {Ouerghi}}, \bibinfo {author} {\bibfnamefont
  {B.}~\bibnamefont {Gil}},  \emph {et~al.},\ }\href {\doibase
  10.1103/PhysRevLett.127.137401} {\bibfield  {journal} {\bibinfo  {journal}
  {Physical Review Letters}\ }\textbf {\bibinfo {volume} {127}},\ \bibinfo
  {pages} {137401} (\bibinfo {year} {2021})}\BibitemShut {NoStop}%
\bibitem [{\citenamefont {Schu{\'e}}\ \emph {et~al.}(2019)\citenamefont
  {Schu{\'e}}, \citenamefont {Sponza}, \citenamefont {Plaud}, \citenamefont
  {Bensalah}, \citenamefont {Watanabe}, \citenamefont {Taniguchi},
  \citenamefont {Ducastelle}, \citenamefont {Loiseau},\ and\ \citenamefont
  {Barjon}}]{schue2019bright}%
  \BibitemOpen
  \bibfield  {author} {\bibinfo {author} {\bibfnamefont {L.}~\bibnamefont
  {Schu{\'e}}}, \bibinfo {author} {\bibfnamefont {L.}~\bibnamefont {Sponza}},
  \bibinfo {author} {\bibfnamefont {A.}~\bibnamefont {Plaud}}, \bibinfo
  {author} {\bibfnamefont {H.}~\bibnamefont {Bensalah}}, \bibinfo {author}
  {\bibfnamefont {K.}~\bibnamefont {Watanabe}}, \bibinfo {author}
  {\bibfnamefont {T.}~\bibnamefont {Taniguchi}}, \bibinfo {author}
  {\bibfnamefont {F.}~\bibnamefont {Ducastelle}}, \bibinfo {author}
  {\bibfnamefont {A.}~\bibnamefont {Loiseau}}, \ and\ \bibinfo {author}
  {\bibfnamefont {J.}~\bibnamefont {Barjon}},\ }\href {\doibase
  10.1103/PhysRevLett.122.067401} {\bibfield  {journal} {\bibinfo  {journal}
  {Physical Review Letters}\ }\textbf {\bibinfo {volume} {122}},\ \bibinfo
  {pages} {067401} (\bibinfo {year} {2019})}\BibitemShut {NoStop}%
\bibitem [{\citenamefont {Solozhenko}\ \emph {et~al.}(1998)\citenamefont
  {Solozhenko}, \citenamefont {Häusermann}, \citenamefont {Mezouar},\ and\
  \citenamefont {Kunz}}]{Solozhenko1998}%
  \BibitemOpen
  \bibfield  {author} {\bibinfo {author} {\bibfnamefont {V.~L.}\ \bibnamefont
  {Solozhenko}}, \bibinfo {author} {\bibfnamefont {D.}~\bibnamefont
  {Häusermann}}, \bibinfo {author} {\bibfnamefont {M.}~\bibnamefont
  {Mezouar}}, \ and\ \bibinfo {author} {\bibfnamefont {M.}~\bibnamefont
  {Kunz}},\ }\href {\doibase 10.1063/1.121186} {\bibfield  {journal} {\bibinfo
  {journal} {Applied Physics Letters}\ }\textbf {\bibinfo {volume} {72}},\
  \bibinfo {pages} {1691} (\bibinfo {year} {1998})}\BibitemShut {NoStop}%
\end{thebibliography}%


\begin{thebibliography}{2}%
\makeatletter
\providecommand \@ifxundefined [1]{%
 \@ifx{#1\undefined}
}%
\providecommand \@ifnum [1]{%
 \ifnum #1\expandafter \@firstoftwo
 \else \expandafter \@secondoftwo
 \fi
}%
\providecommand \@ifx [1]{%
 \ifx #1\expandafter \@firstoftwo
 \else \expandafter \@secondoftwo
 \fi
}%
\providecommand \natexlab [1]{#1}%
\providecommand \enquote  [1]{``#1''}%
\providecommand \bibnamefont  [1]{#1}%
\providecommand \bibfnamefont [1]{#1}%
\providecommand \citenamefont [1]{#1}%
\providecommand \href@noop [0]{\@secondoftwo}%
\providecommand \href [0]{\begingroup \@sanitize@url \@href}%
\providecommand \@href[1]{\@@startlink{#1}\@@href}%
\providecommand \@@href[1]{\endgroup#1\@@endlink}%
\providecommand \@sanitize@url [0]{\catcode `\\12\catcode `\$12\catcode
  `\&12\catcode `\#12\catcode `\^12\catcode `\_12\catcode `\%12\relax}%
\providecommand \@@startlink[1]{}%
\providecommand \@@endlink[0]{}%
\providecommand \url  [0]{\begingroup\@sanitize@url \@url }%
\providecommand \@url [1]{\endgroup\@href {#1}{\urlprefix }}%
\providecommand \urlprefix  [0]{URL }%
\providecommand \Eprint [0]{\href }%
\providecommand \doibase [0]{http://dx.doi.org/}%
\providecommand \selectlanguage [0]{\@gobble}%
\providecommand \bibinfo  [0]{\@secondoftwo}%
\providecommand \bibfield  [0]{\@secondoftwo}%
\providecommand \translation [1]{[#1]}%
\providecommand \BibitemOpen [0]{}%
\providecommand \bibitemStop [0]{}%
\providecommand \bibitemNoStop [0]{.\EOS\space}%
\providecommand \EOS [0]{\spacefactor3000\relax}%
\providecommand \BibitemShut  [1]{\csname bibitem#1\endcsname}%
\let\auto@bib@innerbib\@empty
%</preamble>
\bibitem [{\citenamefont {Solozhenko}\ \emph {et~al.}(1998)\citenamefont
  {Solozhenko}, \citenamefont {Häusermann}, \citenamefont {Mezouar},\ and\
  \citenamefont {Kunz}}]{Solozhenko1998}%
  \BibitemOpen
  \bibfield  {author} {\bibinfo {author} {\bibfnamefont {V.~L.}\ \bibnamefont
  {Solozhenko}}, \bibinfo {author} {\bibfnamefont {D.}~\bibnamefont
  {Häusermann}}, \bibinfo {author} {\bibfnamefont {M.}~\bibnamefont
  {Mezouar}}, \ and\ \bibinfo {author} {\bibfnamefont {M.}~\bibnamefont
  {Kunz}},\ }\href {\doibase 10.1063/1.121186} {\bibfield  {journal} {\bibinfo
  {journal} {Applied Physics Letters}\ }\textbf {\bibinfo {volume} {72}},\
  \bibinfo {pages} {1691} (\bibinfo {year} {1998})}\BibitemShut {NoStop}%
\bibitem [{\citenamefont {Karch}\ and\ \citenamefont
  {Bechstedt}(1997)}]{Karch1997}%
  \BibitemOpen
  \bibfield  {author} {\bibinfo {author} {\bibfnamefont {K.}~\bibnamefont
  {Karch}}\ and\ \bibinfo {author} {\bibfnamefont {F.}~\bibnamefont
  {Bechstedt}},\ }\href {\doibase 10.1103/PhysRevB.56.7404} {\bibfield
  {journal} {\bibinfo  {journal} {Phys. Rev. B}\ }\textbf {\bibinfo {volume}
  {56}},\ \bibinfo {pages} {7404} (\bibinfo {year} {1997})}\BibitemShut
  {NoStop}%
\end{thebibliography}%

\end{document}

\title{Supplemental Materials for the paper entitled\\"Pressure Dependence of Electronic, Vibrational and Optical Properties of wurtzite-Boron Nitride"}% Force line breaks with \\

\author{Martino Silvetti}
\affiliation{\piim}
\affiliation{\etsf}
\author{Claudio Attaccalite}%
\affiliation{\cinam}
\affiliation{\etsf}
\author{Elena Cannuccia}
\affiliation{\piim}
\affiliation{\etsf}
\date{\today}% It is always \today, today,%  but any date may be explicitly specified
\maketitle

%\keywords{Suggested keywords}%Use showkeys class option if keyword display desired

Here we provide additional data on the structural, vibrational, electronic and optical properties of $w$BN.\\

\section{Lattice parameters}
\begin{figure}
	\includegraphics[width=0.45\textwidth]{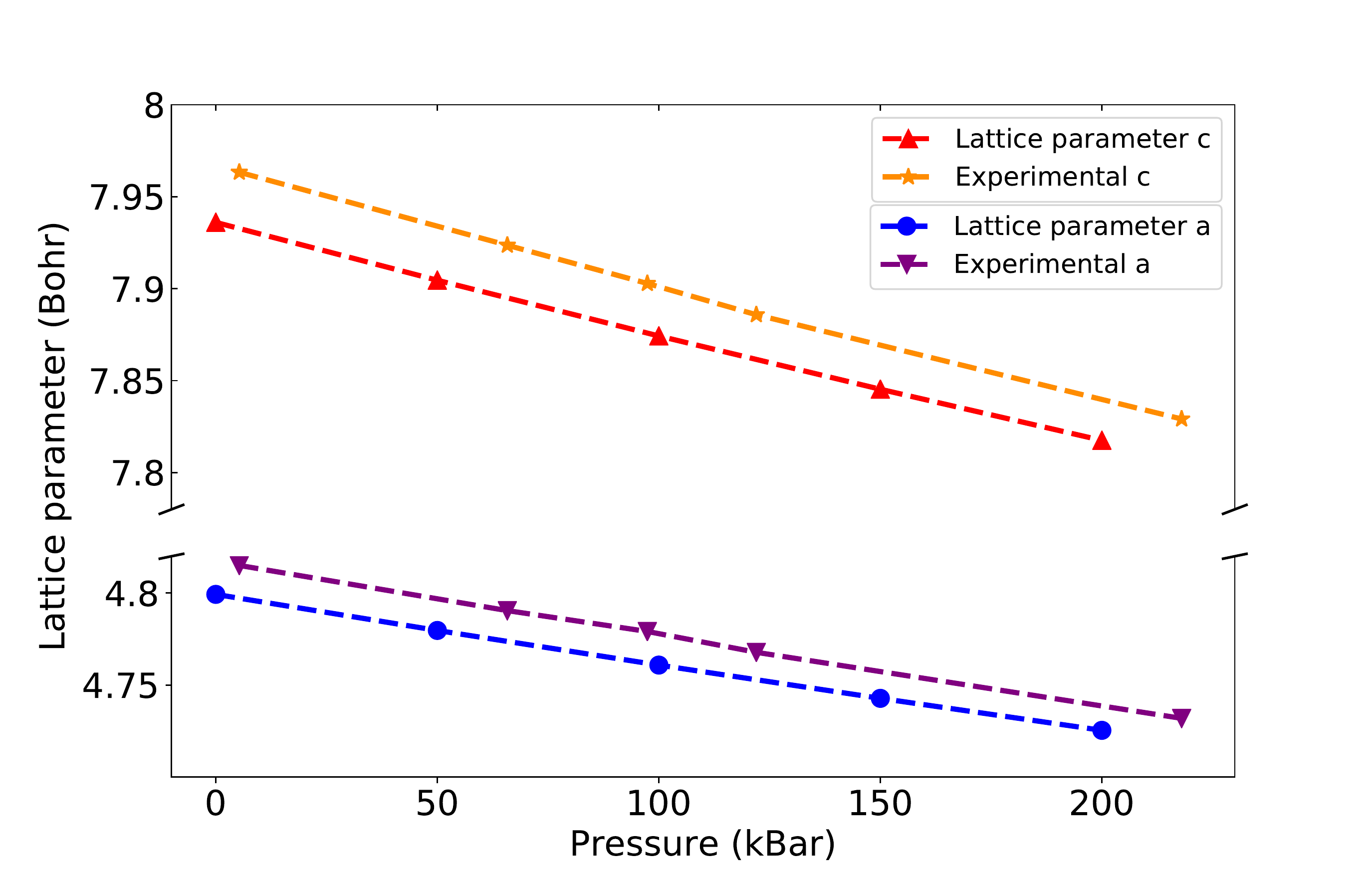}
	\caption{Evolution of DFT optimised lattice paramenters as a function of pressure.}
	\label{fig:Latt_param_vs_P}
\end{figure}

Fig.\,\ref{fig:Latt_param_vs_P} illustrates the variation of lattice parameters $a$ and $c$ as a function of pressure and provides a comparison with
the experimental data\,\cite{Solozhenko1998}.

The corresponding numerical values can be found in Table\,\ref{tab:Latt_param_vs_P}. Our theoretical calculations show a good agreement with the experimental results, with differences ranging from 0.1\% to 0.6\%. 

\begin{table}[h]
	\centering
	\begin{tabular}{c|c|c||c|c|c}
		\hline
		Exp. Pressure  & a  & c & Theo. Pressure & a & c \\
		(kBar)   &  (Bohr)       &   (Bohr)     & (kBar) & (Bohr) & (Bohr)       \\                 
		\hline
		5.3      &  4.82   &   7.96    & 0   & 4.80  &  7.94 \\
		65.8     &  4.79   &   7.92    & 50  & 4.78  &  7.90 \\
		197.4    &  4.78   &   7.90    & 100 & 4.76  &  7.87 \\
		122      &  4.77   &   7.87    & 150 & 4.74  &  7.85 \\
		218      &  4.73   &   7.83    & 200 & 4.73  &  7.82 \\
		\hline 
	\end{tabular}
	\caption{Comparison of experimental and theoretical lattice parameter, with the latter being obtained after relaxation at the given pressure.  }  
	\label{tab:Latt_param_vs_P}
\end{table}

\section{Vibrational properties}
%================================================
\emph{Evolution of LO-TO splitting with pressure}\\
%================================================

In Table\,\ref{tab:S1_LO-TO_press} we report the evolution with pressure of the $E_1$ and $A_1$ LO-TO splitting and the $E_1$-$A_1$ splitting.  
\begin{table}[h]
\centering
\begin{tabular}{c|c|c|c}
\hline
	Pressure  & $E_1$(LO)-$E_1$(TO)  & $A_1$(LO)-$A_1$(TO) & $E_1$(LO)-$A_1$(LO) \\
	 (kBar)   &  ($cm^{-1}$)       &   ($cm^{-1}$)     & ($cm^{-1}$)       \\                 
\hline
	0 & 217.2 & 237.1  & 11.7 \\
       50 & 216.2 & 235.9  & 11.2 \\
      100 & 215.4 & 234.8  & 10.7 \\
      150 & 214.5 & 233.7  & 10.2 \\
      200 & 213.7 & 232.6  & 9.7  \\
\hline 
\end{tabular}
        \caption{Calculated $E_1$, $A_1$ LO-TO splitting and the $E_1$-$A_1$ splitting.}  
	\label{tab:S1_LO-TO_press}
\end{table}

%============================================ BORN EFF CHARGES+DIELE TENSOR ============================================
\emph{Macroscopic dielectric tensor and Born effective charges}\\
%================================================

In $w$BN the dielectric tensor $\ee(\infty)$ is composed of two independent components, one corresponding to the direction parallel to the $c-$axis, $\ee_{\parallel}(\infty)=\ee_{zz}(\infty)$ and the other on the hexagonal plane perpendicular to the $c-$axis, $\ee_{\perp}(\infty)=\ee_{xx}(\infty)=\ee_{yy}(\infty)$.  The average dielectric tensor is defined as $\ee(\infty)=\nicefrac{1}{3}Tr[\ee(\infty)]$. In Table\,\,\ref{tab:S2_DielTens_Z} we report the parallel and perpendicular components of the dielectric tensor at different pressures, which are in agreement with previous results\,\cite{Karch1997}. We observe that, the degree of anisotropy of $\ee(\infty)$ estimated from the difference between $\ee_{\parallel}(\infty)$ and $\ee_{\perp}(\infty)$, $\Delta\ee(\infty)=[\ee_{\parallel}(\infty)-\ee_{\perp}(\infty)]/\ee(\infty)$ does not vary by increasing \emph{hydrostatic} pressure. On the other hand the components of the Born effective charges tensor, the average and the screened effective charges decreases by increasing pressure. 
\begin{table}[h]
\centering
\begin{tabular}{c|c|c|c|c|c|c|c|c}
\hline
	Pressure (kBar)  & $\ee_{\perp}(\infty)$ & $\ee_{\parallel}(\infty)$ & $\ee(\infty)$ & $\Delta\ee(\infty)$ & $Z^{B}_{\perp}$ & $Z^{B}_{\parallel}$ & $Z^{B}$ & $Z^*$     \\
\hline
	0 & 4.52 & 4.677  & 4.57 &  0.033 & 1.832 & 1.928 &  1.864 & 0.871\\
	50 & 4.51 & 4.664  & 4.56 &  0.033 & 1.827 & 1.922 & 1.859 & 0.870\\
	100 & 4.50 & 4.652  & 4.55 &  0.033 & 1.821 & 1.917 & 1.853 & 0.869\\
	150 & 4.49 & 4.641  & 4.54 &  0.033 & 1.816 & 1.911 & 1.848 & 0.867\\
	200 & 4.48 & 4.631  & 4.53 &  0.033 & 1.811 & 1.906 & 1.842 & 0.865\\
\hline 
\end{tabular}
	\caption{Macroscopic dielectric tensor and Born effective charges of $w$BN. }
	\label{tab:S2_DielTens_Z}
\end{table}

%At the K point each mode acquires both transverse and longitudinal characters mixed together and the symmetry is reduced. That results in an elliptic polarization in the oscillations of atoms. For each atom one can identify a \(xy\) 2D reference frame on which the oscillation occurs with amplitudes different for the x and y axis but with equal frequency (\(\omega_{x}=\omega_{y}\)).\\ 

%\st{At the point K no particular symmetry in the phonon mode is identified (QUANTUM ESPRESSO NON TROVA LE RAPPRESENTAZIONI IRR NEL PUNTO K). For all 12 modes an elliptic polarization is introduced which makes atoms oscillate with the same frequency but with different phases that hide the original representation associated with each band. That is a result of the mixing of transverse and longitudinal modes.  It is possible, however, to recognise for each mode at the point K the original irreducible representation associated to it at the point} \(\Gamma\)\st{ by simply walking along the band through the Brillouin zone going from} \(\Gamma\) \st{to K and see how the phases are gradually introduced (QUESTA PARTE} È \st{INUTILE AI FINI DELL'ARTICOLO ED} È \st{COMPLICATA)}. 

%================================
\emph{Phonon modes at point $K$}\\
%================================

\begin{figure}[h]
	\includegraphics[width=0.45\textwidth]{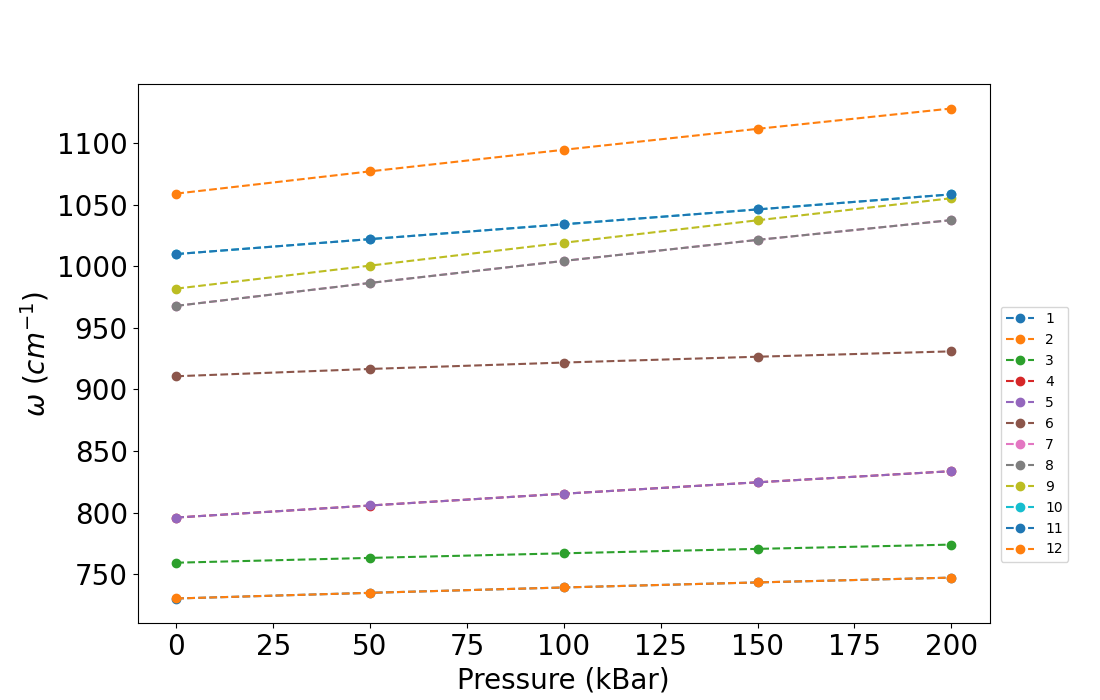}
	\caption{Phonon modes at K point for different pressures.}
	%The key associates to each mode its corresponding irreducible representation at \(\Gamma\)}
	\label{fig:PH_freqKpoint}
\end{figure}

In Fig.~\ref{fig:PH_freqKpoint} we report the evolution of phonon frequencies at point K as a function of pressure and in Table~\ref{tab:phfreq} we resume the phonon frequencies at $0$ kBar. %\MS{Note that at \(K\) the 10\({}^{\mathrm{th}}\) and 11\({}^{\mathrm{th}}\) degenerate modes energy tend to go below the 9\({}^{\mathrm{th}}\) mode energy for pressures larger than 200 kBar.}   \MS{Rimossa tabella dei modi a K, basta il plot \ref{fig:PH_freqKpoint}} 

\begin{table}
	\centering
	%\begin{tabular}{|c|c|c|}
	%	\hline
	%	\# & Symm. & Energy P=0 kBar (cm\({}^{-1}\))\\
	%	\hline
	%	1 - 2   & \(E_{1}\) ac. \(A_{1}\) ac.  & 730.5  \\
	%	3       & \(E_{1}\) ac.                & 759.5  \\
	%	4 - 5   & \(B_{1}^{2}\) \(E_{2}^{1}\)  & 796.16 \\
	%	6       & \(E_{2}^{1}\)                & 910.75 \\
	%	7 - 8   & \(E_{2}^{2}\) \(E_{1}\)      & 967.82 \\
	%	9       & \(E_{2}^{2}\)                & 981.8  \\ 
	%	10 - 11 & \(E_{1}\) \(A_{1}\)          & 1009.83\\
	%	12      & \(B_{1}^{2}\)                & 1058.88 \\
	%	\hline
	%\end{tabular}
	\begin{tabular}{|c|c|}
		\hline
		\# &  Frequency (cm\({}^{-1}\))\\
		\hline
		1 - 2   &  730.5  \\
		3       &  759.5  \\
		4 - 5   &  796.16 \\
		6       &  910.75 \\
		7 - 8   &  967.82 \\
		9       &  981.8  \\ 
		10 - 11 &  1009.83\\
		12      &  1058.88 \\
		\hline
	\end{tabular}
	\caption{Phonon modes at K point at 0 kBar.\label{tab:phfreq}}
	%The key associates to each mode its corresponding irreducible representation at \(\Gamma\)}
\end{table}

%======================================================
\section{DFT Electronic Band Structure and Density of states}
\begin{figure}[h]
	\centering
	\includegraphics[scale=0.3]{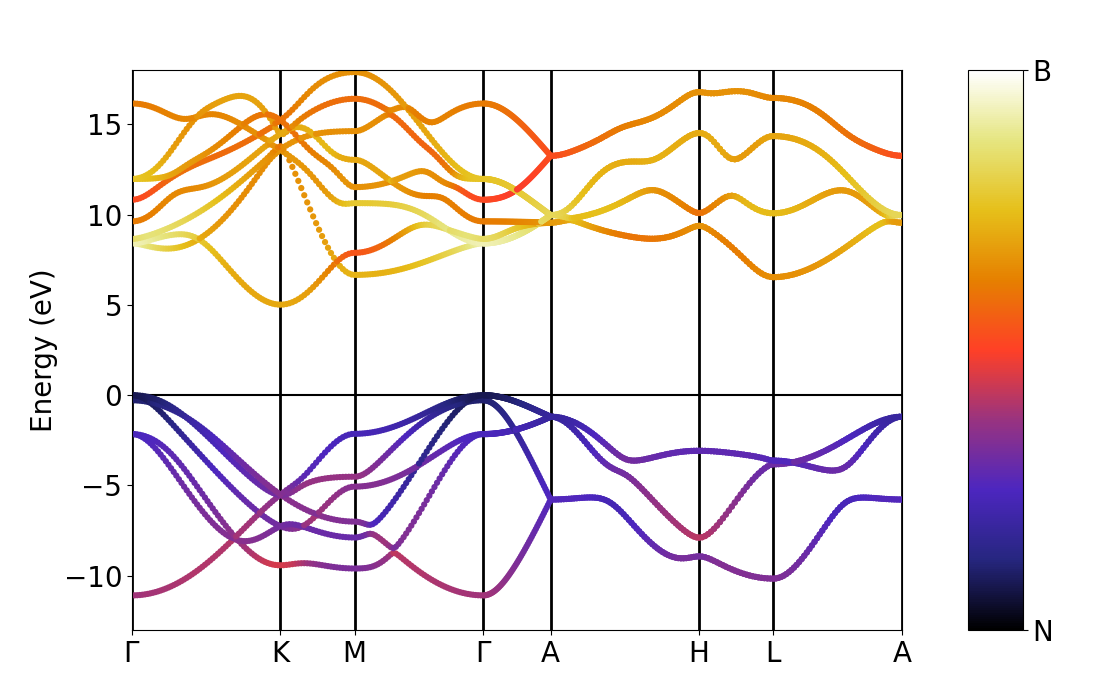}
	\caption{Plot of the DFT band structure of $w$BN at 0 kBar. The colour reveals the weight of Nitrogen and Boron orbitals contribution to each Kohn Sham state.}
	\label{fig:ao_proj_band}
\end{figure}

\begin{figure}[h]
	\centering
	\includegraphics[scale=0.3]{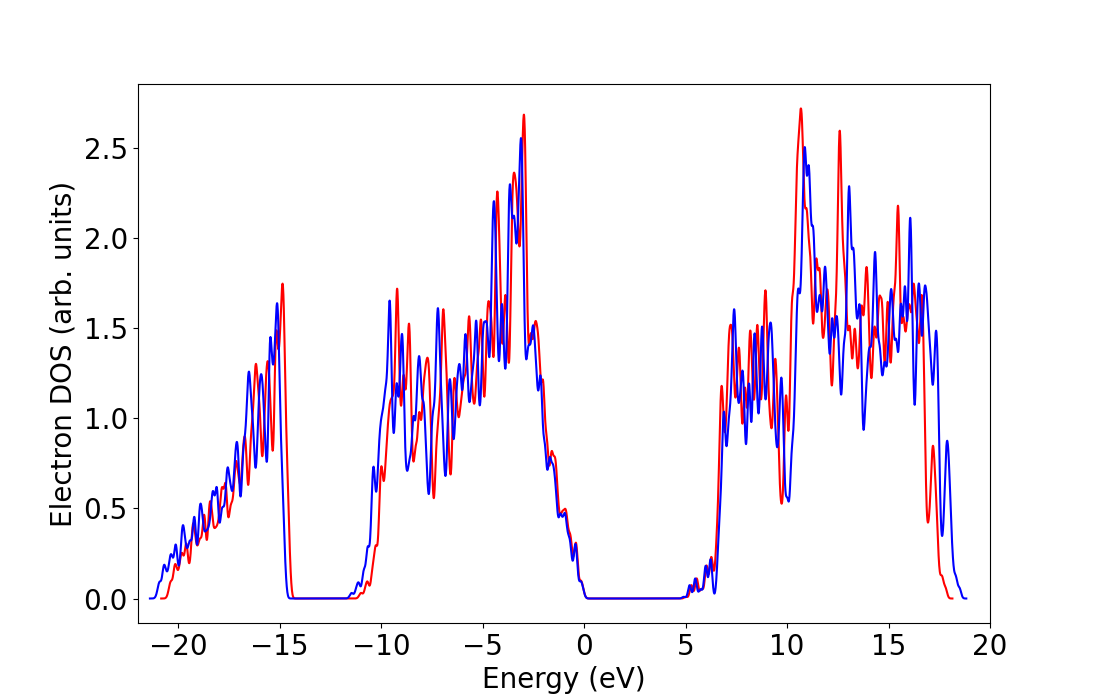}
	\caption{Electron density of state at 0 (red) and 200 kBar (blue). }
	\label{fig:el_dos_tot} 
\end{figure}

\begin{figure}[h]
	\centering
	\includegraphics[scale=0.3]{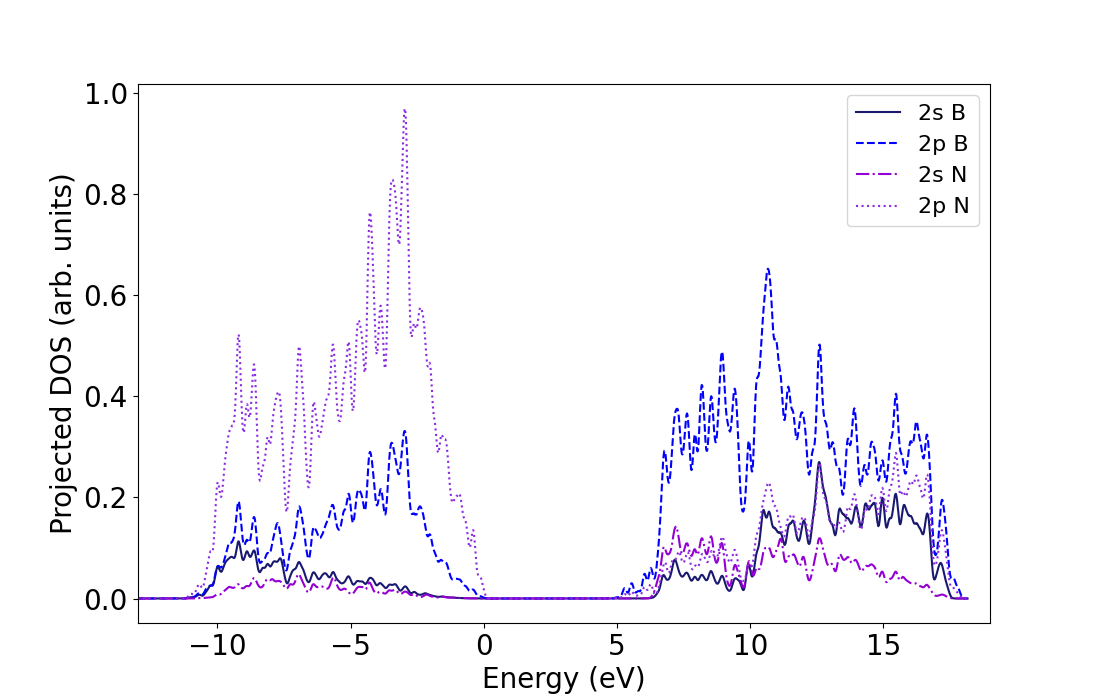}
	\caption{Projected electronic density of state (PDOS) at 0 kBar on 2\(s\) and 2\(p\) orbitals of the B and N atoms of the unit cell. \label{fig:pdos_0kbar}}
\end{figure}

%======================================================
Here we report the $w$BN DFT band structure\,\ref{fig:ao_proj_band}, the density of states\,\ref{fig:el_dos_tot} and the projection on the different atomic orbitals\,\ref{fig:pdos_0kbar}.

%======================================================
%\section{Quasiparticle corrected Electronic Properties}
%======================================================
%The quasi-particle band structure was obtained within $G_0W_0$ approximation\cite{aryasetiawan1998gw}, using a \(\boldsymbol{k}\)-grid $12\times12\times8$ k-points grid, with 75 bands. We used a cutoff of 3 Ha for the dielectric constant that was calculated within the plasmon-pole approximation. 

%We preliminarily converged the number of bands and the size of the response block size (number of $\boldsymbol{G}$ vectors), keeping fixed the . The results we present in our work are obtained with 75 bands and a value of 3 Ha for the response block size. In Fig.~\ref{bandstructQP} we report the band structure obtained with the GW method at 0 kBar.

\section{Light absorption as a function of pressure}
\begin{figure}
   \centering
   \includegraphics[width=0.45\textwidth]{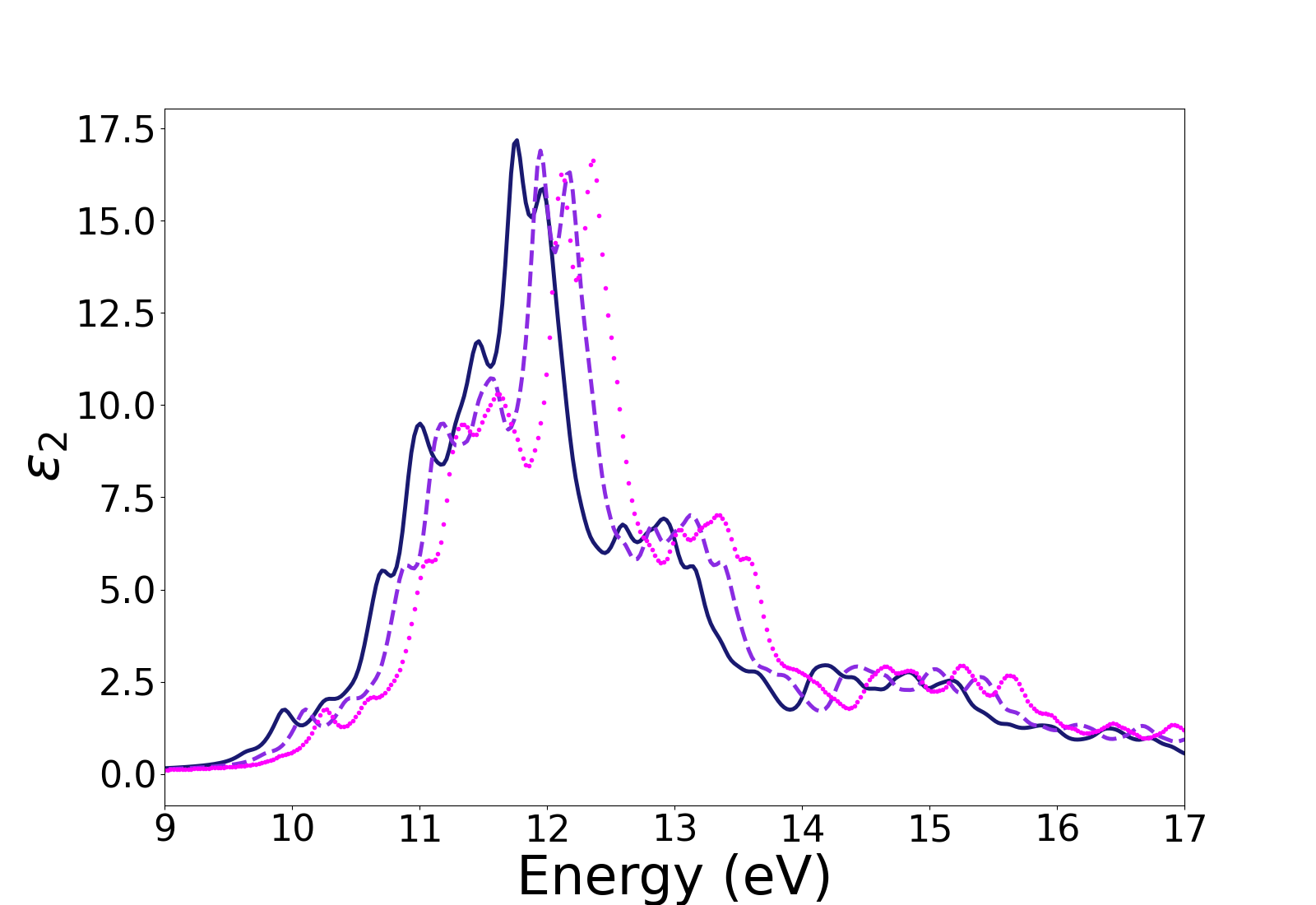}
	\caption{Spectrum for the imaginary part of \(\epsilon_{M}\) at 0 (dark blue, solid line), 100 (violet, dashed line) and 200 kBar (fuchsia, dotted line) obtained from the diagonalization of the BSE Hamiltonian Eq.\,2 of the manuscript.}
	\label{fig:Im_Re_epsilon_bse}
\end{figure}

In Fig.\,\ref{fig:Im_Re_epsilon_bse} we report the calculated imaginary part of the dielectric function. We observe that the absorption spectrum shifts to higher energies as pressure increases reflecting the behaviour of the direct band gap.

\bibliographystyle{apsrev4-1}
\bibliography{wBN}% Produces the bibliography via BibTeX.